\documentclass[12pt]{article}

\font\grande=cmr9.5 scaled \magstep4
\font\medio=cmr9.5 scaled \magstep2
\outer\def\beginsection#1\par{\medbreak\bigskip
      \message{#1}\leftline{\bf#1}\nobreak\medskip
\vskip-\parskip
      \noindent}
\usepackage{graphicx} 
\begin{document}
\bibliographystyle {unsrt}

\titlepage

\begin{flushright}
CERN-PH-TH/2011-126
\end{flushright}

\vspace{10mm}
\begin{center}
{\grande Growth rate of matter perturbations}\\
\vspace{0.5cm}
{\grande as a probe of large-scale magnetism}\\
\vspace{1.5cm}
 Massimo Giovannini\footnote{Electronic address: massimo.giovannini@cern.ch}\\
\vspace{1.5cm}
{{\sl Department of Physics, 
Theory Division, CERN, 1211 Geneva 23, Switzerland }}\\
\vspace{0.5cm}
{{\sl INFN, Section of Milan-Bicocca, 20126 Milan, Italy}}\\
\vspace{0.5cm}
\end{center}

\vskip 0.5cm
\centerline{\medio  Abstract}
The growth rate of matter perturbations is computed in a magnetized environment for the   
 $\Lambda$CDM and $w$CDM paradigms. It is argued that  the baryons do not necessarily follow into the dark matter potential wells after they are released from the drag of the photons. The baryonic evolution equations inherit a forcing term whose explicit form depends on the plasma description and can be deduced, 
for instance, in the resistive magnetohydrodynamical approximation.
After deriving an analytical expression for  the growth rate applicable when dark energy does not cluster, the effects of relativistic corrections and of the inhomogeneities associated with the other species of the plasma are taken into account numerically. The spectral amplitudes and slopes of the stochastic magnetic background are selected to avoid appreciable distortions in the measured temperature and polarization anisotropies of the Cosmic Microwave Background.  The growth of structures in the current 
paradigms of structure formation represents a complementary probe of large-scale magnetism in the same way as the shape of the growth factor and the associated indices can be used, in the conventional lore,  to discriminate between competing scenarios of dark energy or even to distinguish different models of gravity.
\noindent

\vspace{5mm}

\vfill
\newpage
\renewcommand{\theequation}{1.\arabic{equation}}
\setcounter{equation}{0}
\section{Formulation of the problem} 
\label{sec1}
The growth rate of matter inhomogeneities is customarily employed to distinguish the physical features of structure formation scenarios. In terms of  the density 
contrast of the inhomogeneities associated with pressureless matter (i.e. $\delta_{\mathrm{m}}(\vec{x},\tau)$)
 the growth rate will be defined as \cite{peebles}:
\begin{equation}
f(\vec{x},y) = \frac{d \ln{\delta_{\mathrm{m}}}}{d y }, \qquad \delta_{\mathrm{m}}(\vec{x},\tau) = 
\frac{\omega_{\mathrm{c}0}}{\omega_{\mathrm{M}0}} \delta_{\mathrm{c}}(\vec{x},\tau) + 
\frac{\omega_{\mathrm{b}0}}{\omega_{\mathrm{M}0}} \delta_{\mathrm{b}}(\vec{x},\tau),
\label{R1}
\end{equation}
where $y$ will denote, conventionally, the natural logarithm of the scale factor\footnote{The present value of the scale factor is normalized to $1$, i.e. $a_{0}=1$; moreover we will conventionally 
denote $y = \ln{\alpha}$ where $\alpha = a/a_{\mathrm{de}}$ and $a_{\mathrm{de}}$ denotes 
the value of the scale factor at which the critical fractions of matter and dark energy are equal. To avoid potential 
confusion we stress that, in the present script, the notation ``$\ln{}$'' is employed for the natural logarithm while the 
symbol ``$\log{}$" denotes the common logarithm of the corresponding quantity.}, while 
 $\delta_{\mathrm{c}}$ and $\delta_{\mathrm{b}}$ are, respectively, the cold dark matter (CDM) 
density contrast and the baryon density contrast; furthermore, following 
the standard notations, $\omega_{X0} = h_{0}^2 \Omega_{X0}$ where $\Omega_{X0}$ is the (present)
critical fraction of the species $X$ of the plasma. 

The growth rate depends on the dynamical features of the dark energy background and it is affected by relativistic corrections which become relevant
for typical wavelengths of the order of the Hubble radius  \cite{hwang1,hwang2}. 
A rather useful parametrization of the growth rate involves the so-called 
growth index denoted hereunder with $\gamma$:
\begin{equation}
f(y) = \Omega_{\mathrm{M}}^{\gamma}, \qquad  \Omega_{\mathrm{M}} = \frac{\rho_{\mathrm{M}}}{\rho_{\mathrm{t}}},
\label{R2}
\end{equation}
where $\Omega_{\mathrm{M}}= \Omega_{\mathrm{M}}(y)$ is not simply evaluated 
at the present time but it is a $y$-dependent quantity.  
In spite of the homogeneous parametrization of Eq. (\ref{R2}), 
the density contrast of Eq. (\ref{R1}) 
is in general inhomogeneous (see e.g. \cite{hwang1,hwang2}) and its spatial dependence  ranges from wavenumbers 
comparable (and possibly smaller) than the equality wavenumber 
$k_{\mathrm{eq}}$ to wavenumbers directly probed by the currently available large-scale 
structure data, i.e. from $k_{\mathrm{min}} = 0.01 \, h_{0}\, \mathrm{Mpc}^{-1}$ to, approximately, 
$k_{\mathrm{max}} = 0.25 \,h_{0}\, \mathrm{Mpc}^{-1}$. The range $k_{\mathrm{min}} \leq k \leq k_{\mathrm{max}}$ 
includes also the scale at which the spectrum becomes nonlinear, i.e. $k_{\mathrm{nl}} \geq 0.09 \, h_{0} \, 
\mathrm{Mpc}^{-1}$. According to the WMAP 7yr data \cite{wmap7a,wmap7b,wmap7c}, $k_{\mathrm{eq}} = 0.00974_{-0.00040}^{0.00041}\, \mathrm{Mpc}^{-1}$ corresponding to an effective equality multipole $\ell_{\mathrm{eq}} = 137.5\pm 4.3$. Typical wavenumbers  $k \gg k_{\mathrm{eq}}$ crossed inside the Hubble volume before 
matter-radiation equality. Conversely the very large length-scales (relevant for the region of the Sachs-Wolfe (SW)
plateau) fall into the complementary regime $k \ll k_{\mathrm{eq}}$. The SW contribution 
peaks for comoving wavenumbers of the order of $k_{\mathrm{sw}} = 0.0002\, \mathrm{Mpc}^{-1}$  while 
the integrated  SW contribution (typical of the $\Lambda$CDM paradigm and of its extensions) 
peaks between $0.001\,\mathrm{Mpc}^{-1}$ and $0.01\, \mathrm{Mpc}^{-1}$ 
(see, e.g. \cite{isw1,isw2,isw3,isw4,isw5,mg1}). 

The definition\footnote{While different parametrizations of the dark-energy component may lead 
to slightly different semi-analytic determinations of the growth index, these differences will not be essential 
for the purposes of the present investigation.} of $\gamma$ adopted in Eq. (\ref{R2}) (see, e.g. \cite{gamma1,gamma2,gamma3}) implies, in the context of the vanilla 
$\Lambda$CDM scenario, that $ \gamma \simeq 0.55$. 
If the dark-energy component is characterized by a 
constant barotropic ratio $w_{\mathrm{de}} = p_{\mathrm{de}}/\rho_{\mathrm{de}}$ the growth index shall depend on $w_{\mathrm{de}}$. The reference model where $w_{\mathrm{de}}$ is constant is customarily dubbed $w$CDM (see, e.g. \cite{wmap7a,wmap7b,wmap7c}) and this terminology will also be employed in the forthcoming sections of the present paper.

It is not implausible that, in the next decade, dedicated observations will be able to probe  
the growth rate and, more generally, the growth of structures either in the $\Lambda$CDM scenario or in one of its neighboring extensions such as the $w$CDM scenario (see, e.g. \cite{DE1,DE2} for two dedicated reviews on models of dark energy).
An observational scrutiny could even allow for direct cosmological tests of gravity theories as convincingly argued 
in \cite{growth1,growth2}. At the moment the 
large-scale observables directly or indirectly related to the growth rate of matter perturbations are the integrated 
Sachs-Wolfe effect \cite{isw1,isw2,isw3,isw4,isw5,mg1} probed by Cosmic Microwave Background (CMB) experiments and the large-scale galaxy distribution scrutinized, for instance, by 
the Sloan digital survey \cite{sdss1,sdss2,sdss3}. Useful complementary informations can be also obtained from $x$-ray bright clusters \cite{cl1,cl2}, from  Lyman-$\alpha$ power spectra and from weak lensing observations \cite{wl1,wl2}. 

Large-scale magnetic fields can affect the growth rate of matter perturbations. Consequently the determinations of the
 growth index can be used as a potential probe of large-scale magnetism. The aim of the present  investigation is to initiate a more systematic scrutiny of the effects of large-scale magnetism on the growth of structure in the same way as, in recent years, the effects of large-scale magnetic fields on the different observables related to the temperature and polarization anisotropies of the CMB have been analyzed (see \cite{mg1,mg2,mg3,mg4} and references therein). 

According to the WMAP 7yr data \cite{wmap7a,wmap7b,wmap7c} photons decouple at a typical redshift $z_{\mathrm{dec}} = 1088.2 \pm 1.2$. Shortly after, also baryons are freed from the Compton drag of the photons 
at a redshift $z_{\mathrm{drag}} = 1020.3 \pm 1.4$. 
Within a simple magnetohydrodynamical  appproximation, for  $z< z_{\mathrm{drag}}$
the baryon fluid is still strongly coupled and globally neutral while the photons follow, approximately, null geodesics. Both for 
$z> z_{\mathrm{dec}}$ and for $z> z_{\mathrm{drag}}$ the equations governing the evolution of the 
baryon density contrast inherit a direct dependence on the magnetic field. The total density 
contrast $\delta_{\mathrm{m}}$, being the weighted sum of the CDM and of the baryon density contrasts, will also be affected by the large-scale magnetic field.  The growth rate appearing in Eq. (\ref{R1}), the growth index of Eq. (\ref{R2}) will therefore bear the mark of large-scale 
magnetism.  As direct bounds on the parameters of the magnetized $\Lambda$CDM and $w$CDM scenarios 
have been obtained from the analysis of the temperature and polarization anisotropies \cite{mg1,mg2}, it is conceivable that complementary bounds can be deduced from the study of the growth of structures. 
It is nonetheless clear that the impact of large-scale magnetism on the growth of structures must me 
carefully computed beyond the conventional perspective where
 the evolution equation of $\delta_{\mathrm{b}}$ is forced by 
the evolution of $\delta_{\mathrm{c}}$ so that, asymptotically, $\delta_{\mathrm{b}}$ will approach the value of 
$\delta_{\mathrm{c}}$:  this phenomenon is dubbed, at least in the linear regime,  by saying that the 
baryons fall into the dark matter potential wells. This is, in practice, also the situation occurring 
in the $\Lambda$CDM paradigm where the dark energy does not cluster and the density 
contrast of the dark energy vanishes (i.e. $\delta_{\mathrm{de}}=0$). Such a perspective is quantitatively 
and qualitatively challenged by the presence of large-scale magnetic fields.
In the latter situation the evolution of the baryons is the result of the competition of the dark-matter forcing with the 
magnetohydrodynamical Lorentz force. 

The implications of large-scale magnetic fields for the paradigms of structure formation have a long history dating back to the 
contributions of Hoyle \cite{hoyle} and Zeldovich \cite{zel1} (see also \cite{MAN1,MAN2} for two dedicated reviews on large-scale magnetic fields in astrophysics and cosmology).  Peebles (see, e.g. \cite{peebles,bb0})  and Wasserman \cite{bb1} 
argued, within a non-relativistic treatment, that the (comoving) coherence scale of the magnetic field 
should be larger than (but of the order of)  the comoving magnetic Jeans length\footnote{We use here the notations 
employed in the bulk of the paper and $B_{\mathrm{L}}$ denotes the comoving amplitude of the regularized magnetic field measured in nG. Recall that $1\,\mathrm{nG} = 10^{-4} \, \mathrm{nT}= 6.9241\times 10^{-29}\, \mathrm{GeV}^{2}$.} 
\begin{equation}
\lambda_{\mathrm{B\,J}}  = 1.90\times 10^{-2} \, \biggl(\frac{\omega_{\mathrm{b}0}}{0.02258}\biggr)^{-1} \, 
\biggl(\frac{B_{\mathrm{L}}}{\mathrm{nG}}\biggr)\,\, \mathrm{Mpc},
\label{R3}
\end{equation}
if the magnetic field is expected to cause an appreciable contribution to the total matter density contrast after photon decoupling. The point of view implicitly endorsed by \cite{peebles,bb1} was, however,  to seed structure formation 
in a purely baryonic Universe and without the help of any adiabatic or isocurvature initial conditions. The ideas spelled out in \cite{peebles,bb1} have been partially revisited in a pure CDM scenario and in a non relativistic set-up by Coles \cite{coles} without dark energy.  The perspective of the present paper is not to endorse an alternative structure formation paradigm but rather to scrutinize the impact of post-recombination magnetic fields on the current $\Lambda$CDM and $w$CDM scenarios. In doing so it will be essential to profit of the results obtained from the analysis of the 
magnetized CMB anisotropies \cite{mg1,mg2}. To prevent the distortions 
induced by a stochastic magnetic field on the temperature and polarization
anisotropies the parameters of the magnetic power spectra must be in specific ranges which slightly differ in the case of the $\Lambda$CDM and in the case of the 
$w$CDM scenario. In summary the analysis of post-recombination effects 
of the large-scale magnetic fields in the $\Lambda$CDM and $w$CDM paradigms
is still an open problem as the lack of specific predictions for the 
magnetized growth rate shows; one of the purposes of the present paper is to remedy this situation by setting a general framework where these themes can be quantitatively addressed.

The layout  of the present paper can be spelled out as follows. In sec. \ref{sec2} the governing equations of the system will be introduced in the synchronous gauge; the connection of the discussion with other gauge-dependent 
and gauge-invariant treatments will be briefly outlined. 
In sec. \ref{sec3} the computation   
of the growth factor shall be specifically addressed. In section \ref{sec4} the analytical estimates will be corroborated by a specific numerical analysis including the effect of the appropriate relativistic corrections and of  the dark energy fluctuations on the final shape of the growth rate. The final remarks and the concluding perspectives are collected in section \ref{sec5}.

\renewcommand{\theequation}{2.\arabic{equation}}
\setcounter{equation}{0}
\section{Governing equations} 
\label{sec2}
\subsection{$\Lambda$CDM and $w$CDM backgrounds}
The toy models extending and complementing the vanilla $\Lambda$CDM scenario are classified in terms of a set 
of reference paradigms relaxing at least one of the assumptions 
of the conventional $\Lambda$CDM model. For instance, while in the $\Lambda$CDM case 
the dark energy perturbations are totally absent, in the $w$CDM  (where $w$ stands for the barotropic index of the 
dark energy background) the dark energy fluctuations affect, indirectly, the evolution 
of the matter density contrast.  Both in the $\Lambda$CDM and in the $w$CDM models 
the background geometry is described by a spatially flat metric of the type 
$\overline{g}_{\mu\nu}(\tau) = a^2(\tau) \eta_{\mu\nu}$ (where $\eta_{\mu\nu}$ is the Minkowski metric). The  
Friedmann-Lema\^itre equations take their standard form 
\begin{equation}
{\mathcal H}^2 = \frac{8 \pi G a^2}{3} \rho_{\mathrm{t}}, \qquad 
{\mathcal H}^2 - {\mathcal H}' = 4 \pi G a^2 (p_{\mathrm{t}} + \rho_{\mathrm{t}}),\qquad {\mathcal H} = \frac{a'}{a},
\label{FL1}
\end{equation}
where the prime denotes a derivation with respect to the conformal time coordinate $\tau$. 
The explicit form of the total energy density and of the total enthalpy density appearing in Eq. (\ref{FL1}) is given by:
\begin{eqnarray}
&& \rho_{\mathrm{t}} = \rho_{\mathrm{e}} + \rho_{\mathrm{i}} +  \rho_{\mathrm{c}} +
\rho_{\gamma} + \rho_{\nu} + \rho_{\mathrm{de}},
\label{FL2}\\
&& \rho_{\mathrm{t}} + p_{\mathrm{t}} =  \rho_{\mathrm{e}} + \rho_{\mathrm{i}} +\rho_{\mathrm{c}}+ \frac{4}{3} (\rho_{\nu} + \rho_{\gamma}) + (w_{\mathrm{de}} + 1) \rho_{\mathrm{de}};
\label{FL3}
\end{eqnarray}
the subscripts in Eqs. (\ref{FL2}) and (\ref{FL3}) denote, respectively, 
the electrons, the ions, the CDM species, the photons, the neutrinos (assumed massless both in the $\Lambda$CDM and in the 
$w$CDM paradigms) and the dark energy contribution. 
For $z< z_{\mathrm{dec}}$  the baryonic matter density 
$\rho_{\mathrm{b}} = \rho_{\mathrm{e}} + \rho_{\mathrm{i}}$ is customarily introduced where $\rho_{\mathrm{e}} = m_{\mathrm{e}} n_{0}$ and $\rho_{\mathrm{i}}= 
m_{\mathrm{i}} n_{0}$ are, respectively, the electron and ion matter densities. 
The comoving concentrations of the electrons and of the ions (i.e. $n_{0} =
a^3 \tilde{n}_{0}$) coincide because of the electric neutrality of the plasma
\footnote{The plasma is globally neutral $n_{\mathrm{i}} = n_{\mathrm{e}} = n_{0}$ 
and the common value of the electron and ion concentrations can be expressed as 
$n_{\mathrm{0}} = \eta_{\mathrm{b}} n_{\gamma}$ where $n_{\gamma}$ is the 
comoving concentration of photons, $\eta_{\mathrm{b}} =  
6.219 \times 10^{-10} [\omega_{\mathrm{b}0}/(0.02773)]  [T_{\gamma 0}/(2.725\,\mathrm{K})]^{-3}$ and $\omega_{\mathrm{b}0}$ is, as in Eq. (\ref{R1}), the critical fraction of baryonic matter multiplied by 
$h_{0}^2$.}. 
In the $\Lambda$CDM (and $w$CDM) paradigms it is sometimes practical to define further combinations of the energy and matter densities:
\begin{equation}
\rho_{\mathrm{M}} = \rho_{\mathrm{e}} + \rho_{\mathrm{i}} + \rho_{\mathrm{c}} = \rho_{\mathrm{b}} + \rho_{\mathrm{c}}, \qquad 
\rho_{\mathrm{R}} = \rho_{\nu} + \rho_{\gamma},
\label{FL4}
\end{equation}
implying that Eqs. (\ref{FL1}) and (\ref{FL2}) can also be written as\footnote{Notice that since the scale factor
is normalized in such a way that $a_{0}=1$, then ${\mathcal H}_{0} = H_{0} = 100 \, h_{0}\, \mathrm{km}/[
\mathrm{sec} \, \mathrm{Mpc}]$.}:
\begin{eqnarray}
{\mathcal H}^2 &=& {\mathcal H}_{0}^2 \,\biggl(\frac{a_{0}}{a}\biggr)\, \Omega_{\mathrm{M}0}\,\biggl[ 1 + \biggl(\frac{a_{\mathrm{eq}}}{a}\biggr) + \biggl(\frac{a_{\mathrm{de}}}{a}\biggr)^{3 w_{\mathrm{de}}}\biggr],
\label{FL5}\\
{\mathcal Z}(a) &=& \frac{{\mathcal H}'}{{\mathcal H}^2} = - \frac{1}{2} - \frac{3}{2} \biggl[ w_{\mathrm{de}} \, 
( 1 - \Omega_{\mathrm{M}})  + \biggl( \frac{1}{3} - w_{\mathrm{de}}\biggr) \Omega_{\mathrm{R}}\biggr],
\label{FL6}
\end{eqnarray}
where the redshifts of matter radiation equality (i.e. $a_{\mathrm{eq}}$) and dark energy dominance (i.e. 
$a_{\mathrm{de}}$) are defined in terms of the present critical fractions of matter, radiation and dark energy:
\begin{equation}
z_{\mathrm{eq}} + 1 = \frac{a_{0}}{a_{\mathrm{eq}}} = \frac{\Omega_{\mathrm{M}0}}{\Omega_{\mathrm{R}0}}, \qquad
 z_{\mathrm{de}} + 1 = \frac{a_{0}}{a_{\mathrm{de}}} = \biggl(\frac{\Omega_{\mathrm{M}0}}{\Omega_{\mathrm{de}0}}\biggr)^{\frac{1}{3 w_{\mathrm{de}}}}= \biggl( \frac{1}{\Omega_{\mathrm{de}}} -1\biggr)^{\frac{1}{3 w_{\mathrm{de}}}}.
\label{FL7}
\end{equation}
In Eq. (\ref{FL7}) $\Omega_{\mathrm{M}0}$, $\Omega_{\mathrm{R}0}$ and $\Omega_{\mathrm{de}0}$
denote, respectively, the present critical fractions of matter, radiation and dark energy. Conversely, 
in Eqs. (\ref{FL5}) and (\ref{FL6}) the time-dependent critical fractions are defined as:
\begin{eqnarray}
\Omega_{\mathrm{M}}(\alpha) &=& \frac{(z_{\mathrm{eq}} +1)\,\alpha^{3 w_{\mathrm{de}}}}{(z_{\mathrm{eq}} +1) \alpha^{3 w_{\mathrm{de}}}
+ (z_{\mathrm{de}} +1) \alpha^{3 w_{\mathrm{de}}-1}
+ (z_{\mathrm{eq}} +1)},
\nonumber\\
 \Omega_{\mathrm{de}}(\alpha) &=& \frac{(z_{\mathrm{eq}} +1)}{(z_{\mathrm{eq}} +1) \alpha^{3 w_{\mathrm{de}}}+ (z_{\mathrm{de}} +1) 
\alpha^{3 w_{\mathrm{de}}-1} + (z_{\mathrm{eq}} +1)},
\label{FL8}
\end{eqnarray}
where, as in Eq. (\ref{R1}), $\alpha = a/a_{\mathrm{de}}$ 
and $\Omega_{\mathrm{R}}(\alpha) = [1 - \Omega_{\mathrm{M}}(\alpha) - \Omega_{\mathrm{de}}(\alpha)]$ which also 
implies $\alpha \Omega_{\mathrm{M}}(\alpha) =(z_{\mathrm{de}} +1) \Omega_{\mathrm{R}}(\alpha) /(z_{\mathrm{eq}} +1)$.
When $a \gg a_{\mathrm{eq}}$ the contribution of $\Omega_{\mathrm{R}}(\alpha)$ is by definition subleading in comparison 
with $\Omega_{\mathrm{M}}(\alpha)$ and $\Omega_{\mathrm{de}}(\alpha)$. In the limit 
$\Omega_{\mathrm{R}} \ll \Omega_{\mathrm{M}}$,  ${\mathcal Z}(\alpha)$ can be approximated as
\begin{equation}
\lim_{\Omega_{\mathrm{R}} \ll \Omega_{\mathrm{M}}} {\mathcal Z}(\alpha) \to \overline{{\mathcal Z}}(\alpha) = - \frac{1}{2} - \frac{3}{2} w_{\mathrm{de}} ( 1 - \Omega_{\mathrm{M}}).
\label{FL7a}
\end{equation}
By introducing the natural logarithm of the normalized scale factor, i.e. $y = \ln{\alpha}$, it is practical to recall, for future convenience, the following relation
\begin{equation}
\frac{d \Omega_{\mathrm{M}}}{d y} = 3 \Omega_{\mathrm{M}}\biggl[ w_{\mathrm{de}} (1 - \Omega_{\mathrm{M}}) + 
\biggl( \frac{1}{3} - w_{\mathrm{de}}\biggr)\Omega_{\mathrm{R}}\biggr],
\label{FL9}
\end{equation}
which can be used for changing the integration (or derivation) variables from $y$ to $\Omega_{\mathrm{M}}$ itself. In the $\Lambda$CDM model the illustration of the analytical and numerical results 
will be given in terms of the following fiducial set of parameters determined on the basis 
of the WMAP 7yr data alone \cite{wmap7a,wmap7b,wmap7c}:
\begin{equation}
( \Omega_{\mathrm{b}0}, \, \Omega_{\mathrm{c}0}, \Omega_{\mathrm{de}0},\, h_{0},\,n_{\mathrm{s}},\, \epsilon_{\mathrm{re}}) \equiv 
(0.0449,\, 0.222,\, 0.734,\,0.710,\, 0.963,\,0.088),
\label{FL10}
\end{equation}
where, by definition, $\Omega_{\mathrm{M}0} = \Omega_{\mathrm{c}0} + \Omega_{\mathrm{b}0}$; furthermore 
$\epsilon_{\mathrm{re}}$ denotes the optical depth at reionization and $n_{\mathrm{s}}$ is the spectral index 
of curvature perturbations  assigned as\footnote{The specific definition of ${\mathcal R}$ in terms of the variables 
employed in the present paper is reminded hereunder at Eqs. (\ref{gauge4}) and (\ref{gauge5}).}
\begin{equation}
\langle {\mathcal R}(\vec{k},\tau) {\mathcal R}(\vec{p},\tau) \rangle = \frac{2 \pi^2}{k^3} 
{\mathcal P}_{{\mathcal R}}(k) \delta^{(3)}(\vec{k} + \vec{p}), \qquad {\mathcal P}_{{\mathcal R}}(k) = {\mathcal A}_{{\mathcal R}} \biggl(\frac{k}{k_{\mathrm{p}}}\biggr)^{n_{\mathrm{s}}-1}.
\label{FL11}
\end{equation}
For the set of parameters of Eq. (\ref{FL10}) the amplitude of the curvature perturbations at the pivot scale 
$k_{\mathrm{p}}$ is given by  ${\mathcal A}_{{\mathcal R}} = (2.43\pm 0.11)\times 10^{-9}$. It must be stressed 
that the conventions adopted here to assign the power spectra of curvature perturbations coincide 
with the conventions of the WMAP collaboration and will be consistently followed 
hereunder when assigning the power spectra of the fully inhomogeneous magnetic fields (see 
Eq. (\ref{PS2}) and discussion therein). 
If $w_{\mathrm{de}} \neq -1$  the fiducial set of parameters determined from the same observational data 
(but assuming $w$CDM scenario) is given by:
\begin{equation}
( \Omega_{\mathrm{b}0}, \, \Omega_{\mathrm{c}0}, \Omega_{\mathrm{de}0},\, h_{0},\,n_{\mathrm{s}},\, \epsilon_{\mathrm{re}}) \equiv (0.052,\, 0.26,\, 0.67,\,0.69,\, 0.952,\,0.084),
\label{FL12}
\end{equation}
with $w_{\mathrm{de}} = - 1.20^{+0.57}_{-0.58}$ and ${\mathcal A}_{\mathcal R} = (2.50 \pm 0.14) \times 
10^{-9}$. By analyzing different data sets in the light of the $w$CDM scenario,  the error bars on $w_{\mathrm{de}}$ either increase or they are restricted to an interval pinning down the $\Lambda$CDM value $w_{\mathrm{de}} = -1$. In the numerical applications and for illustrative purposes we shall often fix the cosmological 
parameters as in Eq. (\ref{FL10}) by allowing for a variation of $w_{\mathrm{de}}$. Values $w_{\mathrm{de}} < -1$ will 
be excluded for practical reasons since, in these cases, the background may evolve towards a singularity in the future 
\cite{sudden1,sudden2}.
This is, of course, only a practical choice since, in principle the present discussion can also be extended to the 
situation where $w_{\mathrm{de}} < -1$ where interesting models 
may exist \cite{sudden1,sudden2,sudden3,sudden4,sudden5}.

There are three redshift scales which will be used both in the analytical considerations 
as well as in the numerical discussion: the redshift of last scattering (coventionally denoted by $z_{*}$),
the redshift of photon decoupling (conventionally denoted by $z_{\mathrm{dec}}$) and the ``drag" redshift 
 marking the end of photon drag on the baryons (conventionally denoted by $z_{\mathrm{drag}}$).
Following the notations already introduced in Eq. (\ref{R1}) 
the value of $z_{*}$ can be expressed as (see, e.g. \cite{mg2} and references therein):
\begin{eqnarray}
z_{*} &=& 1048[ 1 + (1.24 \times 10^{-3})\, \omega_{\mathrm{b}0}^{- 0.738}] [ 1 + g_{1}\, \omega_{\mathrm{M}0}^{\,\,\,g_2}],
\label{TP6}\\
g_{1} &=& \frac{0.0783 \, \omega_{\mathrm{b}0}^{-0.238}}{[1 + 39.5 \,\,
\omega_{\mathrm{b}0}^{\,\,0.763}]},\qquad 
g_{2} = \frac{0.560}{1 + 21.1 \, \omega_{\mathrm{b}0}^{\,\,1.81}}.
\label{TP7}
\end{eqnarray}
The parameters of Eq. (\ref{FL10}), once inserted into 
Eqs. (\ref{TP6})--(\ref{TP7}) imply 
$z_{*} =1090.77$ in excellent agreement with the direct determination 
obtained in terms of the WMAP 7yr data \cite{wmap7a,wmap7b,wmap7c}
giving $z_{*} = 1090.79^{+0.94}_{-0.92}$.
The redshift of photon decoupling is rather close to $z_{*}$ so that for the accuracy 
of the estimates presented here we will effectively consider 
that $z_{*} \simeq z_{\mathrm{dec}}$. For instance the WMAP 7yr data analyzed 
in terms of the $\Lambda$CDM paradigm imply $z_{\mathrm{dec}} = 1088.2 \pm 1.2$.
The approximate redshift at which photon drag ceases to be effective 
on the baryons can be expressed as \cite{drag1}:
\begin{eqnarray}
z_{\mathrm{drag}} &=& \frac{1291 \, \omega_{\mathrm{M}0}^{0.251}}{1 + 0.659\, 
\omega_{\mathrm{M}0}^{0.828}}[ 1 + b_{1} \omega_{\mathrm{b}0}^{b2}],
\label{TP8}\\
b_{1} &=& 0.313 \, \omega_{\mathrm{M}0}^{-0.419}[ 1 + 0.607 \, \omega_{\mathrm{M}0}^{0.674} ],
\qquad b_{2} = 0.238 \, \omega_{\mathrm{M}0}^{0.223}.
\label{TP9}
\end{eqnarray}
Again, the parameters of Eq. (\ref{FL10}) imply $z_{\mathrm{drag}} = 1020.3$ which is
compatible with the WMAP 7yr data implying $z_{\mathrm{drag}} = 1020.3 \pm 1.2$.
If the evolution of the critical fractions of matter, radiation and dark energy is parametrized in terms of $y = \ln{a/a_{\mathrm{de}}}$, 
the parameters of Eq. (\ref{FL10}) lead, with obvious notations, to the following three  scales $y_{*}= -6.657$, $y_{\mathrm{dec}}=-6.654$ and $y_{\mathrm{drag}}=-6.589$, i.e.
approximately $y_{*} < y_{\mathrm{dec}} < y_{\mathrm{drag}}$.

\subsection{Magnetized fluctuations of the geometry}
The only source of large-scale inhomogeneity of the $\Lambda$CDM  and $w$CDM paradigms resides 
in the curvature perturbations whose magnitude determines primarily the normalization 
of the matter power spectrum.  The introduction of large-scale magnetic fields in the minimal 
$\Lambda$CDM scenario has been previously scrutinized in a number of
analyses either in specific gauges \cite{mg2,mg3,mg4} or even within earlier gauge-invariant approaches \cite{mg5}. The parameters characterizing the magnetized $w$CDM scenario and the ones of the magnetized $\Lambda$CDM scenario have been discussed, respectively, in \cite{mg1} and in \cite{mg2}. 

The synchronous gauge description
is often preferable for the analysis of magnetized perturbations as pointed out long ago \cite{mg4}. 
Moreover, in the problems 
related to the growth of structures, the synchronous gauge leads to an evolution equation 
for the total matter density contrast which is comparatively simpler than in other gauges. 
Consistently with Refs. \cite{mg2,mg3,mg4} the scalar fluctuations of the geometry in the 
synchronous gauge will be expressed, in Fourier space, as:
 \begin{equation}
\delta^{(\mathrm{S})}_{\mathrm{s}} g_{ij}(k, \tau) = a^2(\tau) \biggl[ \hat{k}_{i} \hat{k}_{j}\, h(k,\tau) 
+ 6\, \xi(k,\tau) \,\biggl(\hat{k}_{i} \hat{k}_{j} - \frac{\delta_{ij}}{3} \biggr)\biggr],
\label{gauge1}
\end{equation}
with $\hat{k}_{i} = k_{i}/|\vec{k}|$. Conversely, in
 the conformally Newtonian gauge the perturbed entries of the metric are given by:
\begin{equation}
\delta^{(\mathrm{cn})}_{\mathrm{s}} g_{00}(k, \tau) = 2 \,a^2(\tau) \phi(k,\tau),\qquad 
\delta^{(\mathrm{cn})}_{\mathrm{s}} g_{ij}(k, \tau) = 2 \,a^2(\tau) \psi(k,\tau) \delta_{ij}.
\label{gauge2}
\end{equation}
The parametrizations of Eqs. (\ref{gauge1}) and (\ref{gauge2}) are related 
by the appropriate coordinate transformations, i.e.  
\begin{eqnarray}
\psi(k,\tau) &=& - \xi(k,\tau) + \frac{{\mathcal H}}{2 k^2} [h(k,\tau) + 6 \xi(k,\tau)]',
\nonumber\\
\phi(k,\tau) &=& - \frac{1}{2k^2} \biggl\{[h(k,\tau) + 6 \xi(k,\tau)]'' + {\mathcal H} [h(k,\tau) + 6 \xi(k,\tau)]'\biggr\}.
\label{gauge3}
\end{eqnarray}
The curvature perturbations on comoving 
orthogonal hypersurfaces, i.e. ${\mathcal R}(k,\tau)$, are defined, in the gauge of Eq. (\ref{gauge1}), as:
\begin{equation}
{\mathcal R}(k,\tau) = \xi + \frac{{\mathcal H} \xi'}{{\mathcal H}^2 - {\mathcal H}'} \to {\mathcal R}^{(\mathrm{cn})}(k,\tau)=- \psi - \frac{{\mathcal H} ({\mathcal H} \phi + \psi')}{{\mathcal H}^2 - {\mathcal H}'},
\label{gauge4}
\end{equation}
where the  arrow denotes the resulting expression  of ${\mathcal R}(k,\tau)$ in the conformally Newtonian gauge obtainable  
by shifting metric fluctuations from one coordinate system to the other; the gauge-invariance of ${\mathcal R}(k,\tau)$ 
implies, obviously, that  ${\mathcal R}(k,\tau)={\mathcal R}^{(\mathrm{cn})}(k,\tau)$.
From the last equality in Eq. (\ref{gauge4}) 
it is also apparent that  when ${\mathcal R}' =0$, $\xi(k) = {\mathcal R}(k)$.
In more general terms, by solving Eq. (\ref{gauge4}) in terms of $\xi$,
it can be shown, after integration by part, that:
\begin{equation}
\xi(k,\tau) = {\mathcal R}(k,\tau) - \frac{{\mathcal H}(\tau)}{a(\tau)} \int_{0}^{\tau} \frac{a(\tau_{1})}{{\mathcal H}(\tau_{1})} {\mathcal R}'(k,\tau_{1}) \, 
d\tau_{1},
\label{gauge5}
\end{equation}
where $\tau_{1}$ is an integration variable and where the prime denotes, as usual, a derivation with respect to $\tau$.
Both in analytical and numerical calculations the normalization of the 
curvature perturbations is customarily expressed in terms of ${\mathcal R}$; for this 
reason Eqs. (\ref{gauge4})  and (\ref{gauge5}) turn out 
to be particularly useful in the explicit estimates.  The  same transformations of Eq. (\ref{gauge3}) can be used to 
gauge-transform the governing equations from the conformally Newtonian 
frame to the synchronous coordinate system\footnote{Not only the metric fluctuations will change under coordinate transformations but also the inhomogeneities of the sources. In particular, it can be 
easily shown that $\delta^{(\mathrm{cn})}\rho = 
\delta^{(\mathrm{S})}\rho - \rho' [(h' + 6 \xi')/(2 k^2)]$. These considerations are relevant in connection 
with the last part of section \ref{sec3}.}. As a general comment on the conventions employed in Eqs. (\ref{gauge1}) and (\ref{gauge2}) 
we recall that they are consistent with the ones of \cite{mg1,mg2} and they differ slightly from the ones of Ref. \cite{bertschinger2} (see also \cite{PV1}). Furthermore, in the present approach the curvature perturbations on comoving 
orthogonal hypersurfaces (i.e. ${\mathcal R}$) are often related, in the large-scale limit, to the curvature perturbations on uniform 
density hypersurfaces (see, e.g. \cite{bardeen} and also \cite{hwang1,hwang2}). Finally, general analyses of large-scale 
magnetic fields performed within the covariant formalism are available in the literature (see, e.g. \cite{cov1}) but they are not central to the present discussion.

The Hamiltonian and momentum constraints stemming, respectively, from the $(00)$ and $(0i)$ (perturbed) Einstein equations are given, in real space, by 
\begin{eqnarray}
&& 2 \nabla^2 \xi + {\mathcal H} h' = - 8 \pi G a^2  \biggl[ \delta_{\mathrm{s}} \rho_{\mathrm{f}} + 
\delta_{\mathrm{s}} \rho_{\mathrm{de}} + \delta_{\mathrm{s}} \rho_{\mathrm{B}} + \delta_{\mathrm{s}} \rho_{\mathrm{E}}\biggr],
\label{ham1}\\
&& \nabla^2 \xi'  =  4\pi G a^2 \biggl[ (p_{\mathrm{t}} + \rho_{\mathrm{t}}) \theta_{\mathrm{t}}
+ \frac{\vec{E} \times \vec{B}}{4\pi a^4 } \biggr],
\label{mom1}
\end{eqnarray}
where, $\delta_{\mathrm{s}} \rho_{\mathrm{f}} = \delta_{\mathrm{s}} \rho_{\mathrm{b}} + \delta_{\mathrm{s}} \rho_{\mathrm{c}}+\delta_{\mathrm{s}} \rho_{\nu} + \delta_{\mathrm{s}} \rho_{\gamma}$ 
denotes the density fluctuation of the fluid sources in the synchronous gauge.
In Eq. (\ref{mom1}) the three-divergence of the total velocity field, i.e.   $\theta_{\mathrm{t}} = \vec{\nabla} \cdot\vec{v}_{\mathrm{t}}$ can be expressed in terms of the contribution of each individual species\footnote{
Following exactly the same conventions established in
 Eqs. (\ref{FL2}) and (\ref{FL3}), the various subscripts denote the velocities 
of the different fluid components; recall that, for a generic species $X$, the notation 
$\theta_{X} = \vec{\nabla} \cdot \vec{v}_{X}$.}
\begin{equation}
(p_{\mathrm{t}} + \rho_{\mathrm {t}}) \theta_{\mathrm{t}} = \frac{4}{3} \rho_{\nu} \theta_{\nu} + 
\frac{4}{3} \rho_{\gamma} \theta_{\gamma} +  \rho_{\mathrm{b}} \theta_{\mathrm{b}} + \rho_{\mathrm{c}} \theta_{\mathrm{c}} + 
( w_{\mathrm{de}} +1) \rho_{\mathrm{de}} \theta_{\mathrm{de}}.
\label{mom2}
\end{equation}
In Eqs. (\ref{ham1}) and (\ref{mom1}) the effects of the large-scale electromagnetic fields have been included 
in terms of the comoving electric and magnetic fields 
$\vec{E}(\vec{x},\tau) = a^2(\tau) \vec{{\mathcal E}}(\vec{x},\tau)$ and $\vec{B}(\vec{x},\tau) = a^2(\tau) 
\vec{{\mathcal B}}(\vec{x},\tau)$, i.e. 
\begin{equation}
\delta_{\mathrm{s}} \rho_{\mathrm{B}} = \frac{B^2}{8\pi a^4},\qquad 
\delta_{\mathrm{s}} \rho_{\mathrm{E}} =  \frac{E^2}{8\pi a^4},\qquad 
\delta_{\mathrm{s}} p_{\mathrm{B}} = \frac{\delta_{\mathrm{s}}\rho_{\mathrm{B}}}{3},\qquad \delta_{\mathrm{s}} p_{\mathrm{E}} = \frac{\delta_{\mathrm{s}}\rho_{\mathrm{E}}}{3},
\label{s5}
\end{equation}
where $B^2 = |\vec{B}|^2$ and $E^2 = |\vec{E}|^2$.
For $z > z_{\mathrm{drag}}$ the baryon and photon velocities effectively coincide with $\theta_{\gamma\mathrm{b}}$
\begin{equation}
\theta_{\gamma\mathrm{b}} \simeq \theta_{\gamma} \simeq \theta_{\mathrm{b}}, \qquad \theta_{\mathrm{b}} = 
\frac{m_{\mathrm{e}} \theta_{\mathrm{e}} + m_{\mathrm{i}} \theta_{\mathrm{i}}}{m_{\mathrm{e}} + m_{\mathrm{i}}},
\label{mom4}
\end{equation}
and Eq. (\ref{mom2}) can be written as:
\begin{equation}
(p_{\mathrm{t}} + \rho_{\mathrm{t}}) \theta_{\mathrm{t}} = \frac{4}{3} \rho_{\nu} \theta_{\nu} + 
\frac{4}{3} \rho_{\gamma} ( 1 + R_{\mathrm{b}}) \theta_{\gamma\mathrm{b}}  + \rho_{\mathrm{c}} \theta_{\mathrm{c}} + 
( w_{\mathrm{de}} +1) \rho_{\mathrm{de}} \theta_{\mathrm{de}},\qquad R_{\mathrm{b}}(\alpha)= \frac{3\,\omega_{\mathrm{b}0}\, \alpha}{4 \omega_{\gamma 0} (z_{\mathrm{de}} +1)};
\label{mom2a}
\end{equation}
$R_{\mathrm{b}}(\alpha)$ denotes the well known weighted ratio between the baryonic matter density 
and the photon energy density determining, for instance, the acoustic oscillations 
in the temperature autocorrelations of the CMB.  
The evolution equations for $\vec{E}$ and $\vec{B}$ are: 
\begin{eqnarray}
&&\vec{\nabla}\cdot \vec{B} =0,\qquad  \vec{\nabla}\cdot \vec{E} = 4 \pi 
\rho_{\mathrm{q}},
\label{s1}\\
&& \vec{\nabla} \times \vec{E} + \vec{B}' =0, \qquad 
\vec{\nabla} \times \vec{B} = 4 \pi \vec{J} + \vec{E}'
\label{s2}\\
&& \rho_{\mathrm{q}}'  + \vec{\nabla}\cdot \vec{J} =0, 
\qquad \rho_{\mathrm{q}} = e (n_{\mathrm{i}} - n_{\mathrm{e}}),\qquad 
\vec{J} = e (n_{\mathrm{i}} \vec{v}_{\mathrm{i}} - n_{\mathrm{e}} \vec{v}_{\mathrm{e}}).
\label{s3}
\end{eqnarray}
The total current $\vec{J}$ has been expressed in terms of the two-fluid variables, however, as in the case of flat-space 
magnetohydrodynamics, $\vec{J}$ can be related to the 
electromagnetic fields by means of the generalized Ohm law obtained by subtracting the evolution equations for the ions and for the electrons \cite{mg3}:
\begin{equation}
\vec{J} =\sigma  \biggl(\vec{E} + \vec{v}_{\mathrm{b}} \times \vec{B} + \frac{\vec{\nabla} p_{\mathrm{e}}}{e\,n_{0}}- \frac{\vec{J}\times \vec{B}}{n_{0} e}\biggr), \qquad \sigma =  \frac{\omega_{\mathrm{pe}}^2 }{4\pi\{a \Gamma_{\mathrm{ie}} + (4/3)[\rho_{\gamma}/(n_{0}m_{\mathrm{e}})] \Gamma_{\mathrm{e}\gamma}\}},
\label{s4}
\end{equation}
where $\sigma$ is the conductivity; $\Gamma_{\mathrm{ie}}$ and $\Gamma_{\gamma\mathrm{e}}$ are, respectively, 
the  electron-ion and electron-photon interaction rates. The three terms appearing in the 
Ohm's law are, besides the electric field, the drift term (i.e. 
$\vec{v}_{\mathrm{b}} \times \vec{B}$), the thermoelectric term (containing 
the gradient of the electron pressure\footnote{In Eq. (\ref{s4})  the comoving electron pressure is given by
 $p_{\mathrm{e}} = n_{\mathrm{e}} T_{\mathrm{e}}$ where $n_{\mathrm{e}}$ and $T_{\mathrm{e}}$ are, respectively, the comoving concentration and the comoving temperature of the electrons.}) and the Hall term (i.e. 
$\vec{J}\times \vec{B}$). For frequencies 
much smaller than the (electron) plasma frequency and for typical length-scales much larger than the  Debye screening length the Hall and thermoelectric terms are subleading for the purposes of the present analysis
both before and after photon decoupling \cite{mg2}.

In the resistive magnetohydrodynamical description adopted in the present paper the electric fields 
are subleading in comparison with the magnetic fields by a dimensionless factor going approximately
$(L \, \sigma)^{-1}$ where $L$ is the typical length-scale which can be identified with the wavelength 
of the fluctuation.  For $z< z_{\mathrm{dec}}$ the ionization fraction drops and by $z_{\mathrm{drag}}$ 
the conductivity can be estimated as
\begin{equation}
\sigma = 4.36 \times 10^{-7}\, \mathrm{eV} \biggl(\frac{T_{\gamma 0}}{2.725\, \mathrm{K}}\biggr)^{3/2} \biggl(\frac{\omega_{\mathrm{M}0}}{0.1334}\biggr)^{1/2} \biggl(\frac{z_{\mathrm{drag}} +1}{z_{\mathrm{eq}}+1}\biggr)^{1/2},
\label{con1}
\end{equation}
which is a pretty large value as it can be argued by computing explicitly the suppression parameter arising, for instance, in the magnetic diffusivity terms
\begin{equation}
\frac{k^2}{{\mathcal H} \sigma} \simeq 3 \times  10^{-20} \biggl(\frac{k}{\mathrm{Mpc}}\biggr)^2 \biggl(\frac{z_{\mathrm{drag}}}{1020.3}\biggr)^{-1/2} \, \biggl(\frac{h_{0}}{0.71}\biggr)^{-1},
\label{con2}
\end{equation}
where we estimated $\Omega_{\mathrm{M}}(y_{\mathrm{drag}})= 0.350\Omega_{\mathrm{M}0}$. 
All the terms explicitly suppressed by the conductivity and by one or two 
spatial gradients of the magnetic field will therefore be neglected since they are subleading in comparison with the other terms. 
In spite of the smallness of the ionization fraction after decoupling (and after the drag time) the plasma 
approximation is still very good and the plasma is an excellent 
conductor as Eqs. (\ref{con1})--(\ref{con2}) clearly show.  
When the ionization fraction $x_{\mathrm{e}}$ drops almost suddenly from $1$ to about
$10^{-5} \omega_{\mathrm{M}0}/\omega_{\mathrm{b}0}$ \cite{rec1,rec2} 
(see also \cite{rec3,rec4} for a delayed recombination scenario) 
the concentration of the free 
charge carriers diminishes. Since the free charge carriers 
drop faster than the temperature the Debye scale increases. Overall, however, the plasma 
parameter decreases since 
\begin{equation}
g_{\mathrm{plasma}} = \frac{1}{V_{\mathrm{D}} n_{0} x_{\mathrm{e}}} = 24 e^{3}  \sqrt{\frac{\zeta(3)}{\pi}} 
\sqrt{x_{\mathrm{e}} \eta_{\mathrm{b}0}}
= 2.3\times 10^{-7} \sqrt{x_{\mathrm{e}}} \biggl(\frac{h_{0}^2\Omega_{\mathrm{b0}}}{0.02258}\biggr)^{1/2},
\label{con4}
\end{equation}
where $V_{\mathrm{D}} = 4 \pi \lambda_{\mathrm{D}}^3/3$ is the volume of the Debye sphere, 
$\lambda_{\mathrm{D}}$ is the Debye length and $\zeta(3)=1.202$. Equation (\ref{con4})  shows that the plasma approximation 
is still rather accurate for $z< z_{\mathrm{drag}}$; furthermore the largeness of the conductivity 
justifies the approximations adopted in the present analysis. At least in the general equations written below
the terms involving the electric fields will be kept both for sake of completeness and for future convenience.

The electromagnetic pressure and all the sources of anisotropic stress enter the perturbed $(ij)$ components of the Einstein equations:
\begin{eqnarray}
&& h'' + 2 {\mathcal H} h' + 2 \nabla^2 \xi = 24 \pi G a^2 [ \delta_{\mathrm{s}} p_{\mathrm{f}} + \delta_{\mathrm{s}} p_{\mathrm{de}} + \delta_{\mathrm{s}} p_{\mathrm{B}} + \delta p_{\mathrm{E}}], 
\label{ij1}\\
&& \frac{1}{a^2} [a^2 \,(h + 6 \xi)' \,]' + 2 \nabla^2 \xi = 32 \pi G  a^2 \biggl[ \rho_{\nu} \sigma_{\nu} + \rho_{\gamma} \sigma_{\mathrm{B}}
+ \rho_{\gamma} \sigma_{\mathrm{E}}  
+ \frac{3}{4} \rho_{\mathrm{de}} (w_{\mathrm{de}} +1) \sigma_{\mathrm{de}}\biggr],
\label{ij2}
\end{eqnarray}
where $\delta_{\mathrm{s}} p_{\mathrm{f}} = (\delta_{\mathrm{s}} \rho_{\gamma} + \delta_{\mathrm{s}} \rho_{\nu})/3$,
in analogy with $\delta_{\mathrm{s}} \rho_{\mathrm{f}}$, denotes the fluctuation of the pressure of the fluid components. In Eq. (\ref{ij2}) the  notation
\begin{equation}
\partial_{i}\partial_{j} \Pi^{ij}_{\mathrm{t}} = \frac{4}{3} \rho_{\nu} \nabla^2 \sigma_{\nu} + \frac{4}{3} \rho_{\gamma} 
\nabla^2 \sigma_{\mathrm{B}} + \frac{4}{3} \rho_{\gamma} \nabla^2\sigma_{\mathrm{E}} +  \rho_{\mathrm{de}} (w_{\mathrm{de}} +1) \nabla^2 \sigma_{\mathrm{de}},
\label{anis2}
\end{equation}
has been adopted where, as  usual, the total anisotropic stress $\Pi_{\mathrm{t}\,i}^{j}$ 
has been made explicit in terms of the anisotropic stresses of the different species: 
\begin{equation}
\Pi_{i\,\mathrm{t}}^{j} = \Pi_{i\,\nu}^{j}  +  \Pi_{i\,\mathrm{B}}^{j}  + \Pi_{i\,\mathrm{E}}^{j} + \Pi_{i\,\mathrm{de}}^{j}.
\label{anis1}
\end{equation}
The various subscripts in Eq. (\ref{anis1}) denote the corresponding components and, in 
particular the electromagnetic contribution:
\begin{equation}
  \Pi_{i\,\mathrm{B}}^{j} = \frac{1}{4 \pi a^4} \biggl[ B_{i} B^{j} - \frac{\delta_{i}^{j}}{3} B^2 \biggr],\qquad \Pi_{i\,\mathrm{E}}^{j} = \frac{1}{4 \pi a^4} \biggl[ E_{i} E^{j} - \frac{\delta_{i}^{j}}{3} E^2 \biggr].
\label{anisBE}
\end{equation}
The species of the plasma either interact strongly 
with the plasma (like the elctrons, the ions and the photons) 
or they only feel the effects of the geometry (like the CDM 
component and the dark-energy). The baryon evolution equations are given by:
\begin{equation}
\delta_{\mathrm{b}}' = \frac{h'}{2} - \theta_{\mathrm{b}},\qquad 
\theta_{\mathrm{b}}' + {\mathcal H} \theta_{\mathrm{b}}' = \frac{\vec{\nabla} \cdot [ \vec{J} \times \vec{B}]}{a^4 \rho_{\mathrm{b}}} +  \frac{4}{3} \frac{\rho_{\gamma}}{\rho_{\mathrm{b}}} \epsilon' (\theta_{\gamma} - \theta_{\mathrm{b}}),
\label{BP1}
\end{equation}
while the evolution equations for the photons are:
\begin{equation}
\theta_{\gamma}' = - \frac{1}{4} \nabla^2 \delta_{\gamma}+ 
\epsilon' (\theta_{\mathrm{b}} - \theta_{\gamma}),\qquad \delta_{\gamma}' = \frac{2}{3} h'  - \frac{4}{3} \theta_{\gamma}.
\label{BP4}
\end{equation}
In Eq. (\ref{BP4})  $\epsilon' = a \tilde{n}_{0} \sigma_{\mathrm{e}\gamma}$ is the differential optical depth due to electron-photon scattering.
The CDM density contrast and velocity will evolve, respectively, as 
\begin{equation}
\delta_{\mathrm{c}}' = \frac{h'}{2} - \theta_{\mathrm{c}},\qquad \theta_{\mathrm{c}}' + {\mathcal H} \theta_{\mathrm{c}} = 0,
\label{C2}
\end{equation}
while the evolution of the neutrinos is given by: 
\begin{eqnarray}
&&\theta_{\nu}' = - \frac{1}{4} \nabla^2 \delta_{\nu} + \nabla^2 \sigma_{\nu},\qquad \delta_{\nu}' = \frac{2}{3} h' - \frac{4}{3} \theta_{\nu}, 
\label{N2}\\
&& \sigma_{\nu}' = \frac{4}{15} \theta_{\nu} - \frac{3}{10} {\mathcal F}_{\nu 3} - \frac{2}{15} h' - \frac{4}{5} \xi' ,
\label{N3}
\end{eqnarray}
where ${\mathcal F}_{\nu 3}$ reminds of the coupling of the monopole and of the dipole 
to the higher multipoles of the neutrino phase space distribution. The latter term will 
be set to zero in the class of initial conditions discussed in the present paper but it can be relevant 
when magnetized non-adiabatic modes are consistently included.

As far as the dark-energy fluctuations are concerned, the situation changes 
qualitatively between the $\Lambda$CDM scenario and the $w$CDM case. 
If $w_{\mathrm{de}}=-1$, $\delta_{\mathrm{de}} = \theta_{\mathrm{de}} =0$ and the dark
energy does not cluster on subhorizon scales.  If $w_{\mathrm{de}} \neq -1$ 
 the dark energy perturbations evolve; to prevent instabilities the barotropic index $w_{\mathrm{de}}$ and the sound speed $c_{\mathrm{de}}$ are assigned indipendently. The latter choice implies 
that the total pressure fluctuation inherits a non-adiabatic contribution which is proportional 
to $(w_{\mathrm{de}} - c_{\mathrm{de}}^2)$; thus the fluctuations of the 
dark energy pressure can be written as \cite{cs1} (see also \cite{cs2,cs3}):
\begin{equation}
\delta_{\mathrm{s}} p_{\mathrm{de}} = c_{\mathrm{de}}^2  \delta \rho_{\mathrm{de}} + \delta p_{\mathrm{nad}}, \qquad 
\delta p_{\mathrm{nad}} = 3 {\mathcal H} ( 1 + w_{\mathrm{de}}) (  c_{\mathrm{de}}^2 - w_{\mathrm{de}}) 
\rho_{\mathrm{de}} \frac{\theta_{\mathrm{de}}}{k^2},
\label{DE1}
\end{equation}
from which it is clear that $c_{\mathrm{de}}^2$ is the sound speed in the frame comoving 
with the dark energy fluid. Using Eq. (\ref{DE1}) together with the (perturbed) 
covariant conservation equation for the energy-momentum tensor of the dark energy 
the following pair of equations can be readily obtained in Fourier space:
\begin{eqnarray}
&& \delta_{\mathrm{de}}' + 3 {\mathcal H}( c_{\mathrm{de}}^2 - w_{\mathrm{de}}) \delta_{\mathrm{de}} + 
(w_{\mathrm{de}} + 1)\biggl\{ \biggl[k^2 
+ 9 {\mathcal H}^2( c_{\mathrm{de}}^2 - w_{\mathrm{de}})\biggr] \frac{\theta_{\mathrm{de}}}{k^2}  - \frac{h'}{2} \biggr\}=0,
\label{DE2}\\
&& \theta_{\mathrm{de}}' + {\mathcal H} ( 1 - 3 c_{\mathrm{de}}^2) \theta_{\mathrm{de}} 
- \frac{c_{\mathrm{de}}^2 k^2 \delta_{\mathrm{de}}}{(w_{\mathrm{de}} +1) } =0,
\label{DE3}
\end{eqnarray}
In Eq. (\ref{DE2}) the possible presence of an anisotropic stress for the dark energy contribution  has been neglected. 
The bounds on $c_{\mathrm{de}}^2$ are currently rather loose and we shall assume, as customarily 
done, that $0\leq c_{\mathrm{de}}^2 \leq 1$.
The expression reported in Eq. (\ref{DE1}) is mathematically analog to (but physically different from)
the scalar fluctuation of the total fluid pressure which can be written as 
\begin{equation}
\delta_{\mathrm{s}} p_{\mathrm{f}} = c_{\mathrm{s\,t}}^2 \delta_{\mathrm{s}} \rho_{\mathrm{f}} + \delta \overline{p}_{\mathrm{nad}},\qquad c_{\mathrm{s\,t}}^2 = \frac{p_{\mathrm{t}}'}{\rho_{\mathrm{t}}'},
\label{DE4}
\end{equation}
where $ c_{\mathrm{s\,t}}^2$ is now the total sound speed of the ordinary fluid sources (i.e. characterized by a barotropic index which is positive semidefinite) and 
$\delta \overline{p}_{\mathrm{nad}}$ parametrizes the non-adiabatic pressure fluctuations 
stemming from the spatial variation of the chemical composition of the plasma. In the standard case 
$\delta\overline{p}_{\mathrm{nad}} \neq 0$ for the CDM-radiation mode, for the neutrino-density mode, for the 
neutrino velocity mode and for the baryon-radiation mode. All these modes can be 
appropriately generalized to include the contribution of fully inhomogeneous magnetic fields \cite{mg6}, however,
for the present purposes we shall avoid the technical complication of the non-adiabatic 
modes and assume throughout adiabatic initial conditions (i.e. $\delta \overline{p}_{\mathrm{nad}} =0$) 
both at the level of the Einstein-Boltzmann hierarchy and at the level of the growth equation.

\renewcommand{\theequation}{3.\arabic{equation}}
\setcounter{equation}{0}
\section{Estimates of the magnetized growth rate} 
\label{sec3}
The evolution equation for the growth rate introduced in Eq. 
(\ref{R1}) will now be derived and solved in different approximations.
 The combination of Eqs. (\ref{BP1}) and (\ref{C2}) leads to the following 
equation for the matter density contrast:
\begin{equation}
\delta_{\mathrm{m}}'' + {\mathcal H} \delta_{\mathrm{m}}' = - \frac{\omega_{\mathrm{b}0}}{\omega_{\mathrm{M}0}} \frac{\vec{\nabla} \cdot 
( \vec{J} \times \vec{B})}{a^4 \, \rho_{\mathrm{b}}} + \frac{1}{2} (h'' + {\mathcal H} h').
\label{GR1}
\end{equation}
In (\ref{GR1}) no approximations have been made on the relative weight of the relativistic corrections.
By summing up Eqs. (\ref{ham1}) and (\ref{ij1}) the following equation can be obtained
\begin{equation}
h'' + {\mathcal H} h' = 8\pi G a^2 \biggl[ \delta_{\mathrm{s}} \rho_{\mathrm{f}} + 3 \delta_{\mathrm{s}} p_{\mathrm{f}} + \delta_{\mathrm{s}} \rho_{\mathrm{de}} + 3 \delta_{\mathrm{s}} p_{\mathrm{de}}+ \delta_{\mathrm{s}} \rho_{\mathrm{B}} + 3 \delta_{\mathrm{s}} p_{\mathrm{B}}\biggr].
\label{GR2}
\end{equation}
The pressure fluctuations of the fluid sources have been separated, according to Eq. (\ref{DE4}), into  the adiabatic and non-adiabatic contributions; the non-adiabatic 
contribution of the fluid sources will be set to zero and, therefore, 
inserting Eq. (\ref{GR2}) into Eq. (\ref{GR1}), the result is:
\begin{eqnarray}
\delta_{\mathrm{m}}'' + {\mathcal H} \delta_{\mathrm{m}}' &=& - \frac{\omega_{\mathrm{b}0}}{\omega_{\mathrm{M}0}} \frac{\vec{\nabla} \cdot 
( \vec{J} \times \vec{B})}{a^4 \, \rho_{\mathrm{b}}}+ \frac{3}{2} {\mathcal H}^2 \Omega_{\mathrm{M}} \biggl\{ \delta_{\mathrm{m}} + 2 R_{\gamma} \frac{\Omega_{\mathrm{R}}}{\Omega_{\mathrm{M}}} \Omega_{\mathrm{B}}
+ \frac{\Omega_{\mathrm{R}}}{\Omega_{\mathrm{M}}} \delta_{\mathrm{R}} 
\nonumber\\
&+& \frac{\Omega_{\mathrm{de}}}{\Omega_{\mathrm{M}}} \biggl[ 
\delta_{\mathrm{de}}( 1 + 3 c_{\mathrm{de}}^2) + 9 {\mathcal H} ( 1 + w_{\mathrm{de}}) (
c_{\mathrm{de}}^2 - w_{\mathrm{de}}) \frac{\theta_{\mathrm{de}}}{k^2}\biggr] \biggr\}.
\label{GR5}
\end{eqnarray}
 At the right hand side of Eq. (\ref{GR5}) the magnetic energy density 
is expressed in units of the photon energy density by defining
 $\delta_{\mathrm{s}} \rho_{\mathrm{B}} = \Omega_{\mathrm{B}}(\vec{x},\tau) \rho_{\gamma}$. Equation (\ref{GR5}) can be further modified 
by recalling that, on the basis of simple vector identities (see first  paper in \cite{mg6}) 
\begin{equation}
\frac{3}{4} \frac{\vec{\nabla} \cdot 
( \vec{J} \times \vec{B})}{a^4 \, \rho_{\gamma}} =\frac{\partial_{i} \partial_{j} \Pi^{ij}_{\mathrm{B}}}{p_{\gamma} + 
\rho_{\gamma}} - \frac{\nabla^2 \Omega_{\mathrm{B}}}{4}.
\label{GR6}
\end{equation}
For $a\gg a_{\mathrm{eq}}$,  $\Omega_{\mathrm{R}} \ll \Omega_{\mathrm{M}}$ and $\Omega_{\mathrm{R}} \ll \Omega_{\mathrm{de}}$ in Eq. (\ref{GR5}). 
Moreover, assuming $|\delta_{\mathrm{de}}| \ll |\delta_{\mathrm{m}}|$  the approximate form of Eq. (\ref{GR5}) becomes:
\begin{equation}
 \delta_{\mathrm{m}}'' + {\mathcal H} \delta_{\mathrm{m}}'  - \frac{3}{2} {\mathcal H}^2 \Omega_{\mathrm{M}} \delta_{\mathrm{m}} =
 - \frac{\omega_{\mathrm{b}0}}{4\, \omega_{\mathrm{M}0}\, R_{\mathrm{b}} }[ 4 \nabla^2 \sigma_{\mathrm{B}} - \nabla^2 \Omega_{\mathrm{B}}].
 \label{GR7}
\end{equation}
Equation (\ref{GR7}) only applies for  redshifts $z < z_{\mathrm{drag}}$ and will now be used 
for analytic estimates.  In the limit $\Omega_{\mathrm{B}}\to 0$ and 
$\sigma_{\mathrm{B}} \to 0$ when Eq. (\ref{GR7}) reproduces the standard equation 
analyzed in various situations for the calculation of the growth index in diverse models of dark energy 
\cite{gamma1,gamma2,gamma3} (see also \cite{et1,et2,et3,et4,et5}).  
By changing the variable from the conformal time coordinate to the natural logarithm of the normalized 
scale factor, Eq. (\ref{GR7}) becomes, for $y_{\mathrm{i}} > y_{\mathrm{drag}}$,
\begin{equation}
\frac{d^2 \delta_{\mathrm{m}}}{d y^2} + \biggl[ 1 + \overline{{\mathcal Z}}(y) \biggr] \frac{d \delta_{\mathrm{m}}}{d y} - \frac{3}{2} \Omega_{\mathrm{M}} \delta_{\mathrm{m}} = -   \frac{\omega_{\mathrm{b}0}}{4\, \omega_{\mathrm{M}0}\, R_{\mathrm{b}}(y) {\mathcal H}^2(y) }[ 4 \nabla^2 \sigma_{\mathrm{B}} - \nabla^2 \Omega_{\mathrm{B}}],
\label{GR10}
\end{equation}
where, recalling the expressions of Eqs. (\ref{FL5}) and (\ref{mom2a}), 
\begin{eqnarray}
&& R_{\mathrm{b}}(y) = \frac{3}{4} \frac{\omega_{\mathrm{b}0}}{\omega_{\gamma 0}} \frac{e^{y}}{ z_{\mathrm{de}} +1},
\nonumber\\
&& {\mathcal H}^2(y) = {\mathcal H}_{0}^2 \frac{z_{\mathrm{de}} +1}{z_{\mathrm{eq}} +1} \Omega_{\mathrm{M}0}
e^{-y} \biggl[ ( z_{\mathrm{eq}} +1) + ( z_{\mathrm{de}} +1) e^{-y} + (z_{\mathrm{eq}} + 1) e^{- 3 w_{\mathrm{de}} y}\biggr].
\label{GR1010}
\end{eqnarray}
 Since the magnetic fields are stochastically distributed,  the ensemble average of their Fourier modes obeys:
\begin{equation}
\langle B_{i}(\vec{k}) \, B_{j}(\vec{p}) \rangle = \frac{2\pi^2 }{k^3} P_{ij}(k) {\mathcal P}_{\mathrm{B}}(k) \delta^{(3)}(\vec{k} + \vec{p}), \qquad {\mathcal P}_{{\mathrm{B}}}(k) = {\mathcal A}_{\mathrm{B}} \biggl(\frac{k}{k_{\mathrm{L}}}\biggr)^{n_{\mathrm{B}}-1}, 
\label{PS2}
\end{equation}
where $P_{ij}(k) = (k^2 \delta_{ij} - k_{i} k_{j})/k^2$; 
${\mathcal A}_{\mathrm{B}}$ the spectral amplitude of the magnetic field at the pivot scale $k_{\mathrm{L}} = 1\, \mathrm{Mpc}^{-1}$.  The conventions adopted in Eq. (\ref{PS2}) are the same as the ones employed in Eq. (\ref{FL11}) 
reproduces the conventions 
followed by the WMAP collaboration (see, e.g. \cite{wmap7a,wmap7b,wmap7c} and earlier data releases).
There are some who assign the power spectra of curvature perturbations as in Eq. (\ref{FL11}) (where the scale-invariant limit is  $n_{\mathrm{s}} \to 1$) while, on the contrary, the power spectra of magnetic fields are assigned in such a way that their scale-invariant limit would correspond to $n_{\mathrm{B}} \to - 3$ (and not to $n_{\mathrm{B}} \to 1$). It seems deliberately confusing to use different definitions for the same mathematical object in the same physical framework. In other words, denoting with $n$ a generic spectral index,  
it is certainly possible to assign the power spectra appearing in Eqs. (\ref{FL11}) and (\ref{PS2})  by engineering the scale-invariant limit  for  $n \to -3$ rather than for
 $n \to 1$. However this choice must be consistently implemented for all the power spectra.  
In the present paper the power spectra of Eqs. (\ref{FL11}) and (\ref{PS2}) 
are then assigned with the same conventions. Consequently, in Eq. (\ref{PS2}), 
${\mathcal A}_{\mathrm{B}}$ has the correct dimensions of an energy density\footnote{In other words, as observed long ago in a closely related context \cite{mgs}, the magnetic power 
spectra assigned here coincide with the magnetic energy density per logarithmic 
interval of frequency.}. 
Furthermore ${\mathcal A}_{\mathrm{B}}$ 
can be related to the regularized magnetic field intensity $B_{\mathrm{L}}$ \cite{mg1,mg2} which is customarily 
employed to phrase the comoving values of the magnetic filed intensity.
 In the case when  $n_{\mathrm{B}} > 1$ (i.e. blue magnetic field spectra),  ${\mathcal A}_{\mathrm{B}} = 
(2\pi)^{n_{\mathrm{B}} -1} \, B_{\mathrm{L}}^2 /\Gamma[(n_{\mathrm{B}} -1)/2]$; if $n_{\mathrm{B}} < 1$ 
(i.e. red magnetic field spectra), ${\mathcal A}_{\mathrm{B}} =[ (1 -n_{\mathrm{B}})/2] (k_{\mathrm{p}}/k_{\mathrm{L}})^{(1 - n_{\mathrm{B}})}B_{\mathrm{L}}^2$. In the case of white spectra (i.e. $n_{\mathrm{B}} =1$) the two-point function is logarithmically divergent in real space and this is fully analog to what happens in Eq. (\ref{FL11}) when $n_{\mathrm{s}} =1$, i.e. the Harrison-Zeldovich (scale-invariant) spectrum. 

The parameter space  of the magnetized $w$CDM models and of the magnetized  $\Lambda$CDM scenario 
have been investigated, respectively, in \cite{mg1} and \cite{mg2}. In a frequentist approach, the boundaries of the confidence regions obtained in \cite{mg1,mg2} represent exclusion plots at  $68.3\,$\% and $95.4\,$\% confidence level.  
When moving from the magnetized $\Lambda$CDM 
scenario to the magnetized $w$CDM model we have that the 
 the parameters maximizing the likelihood get shifted to slightly larger values\footnote{The difference between Eqs. (\ref{NTT}) and (\ref{NTE}) is that the parameters 
of Eq. (\ref{NTT}) are obtained from the analysis of the temperature autocorrelations while the parameters of Eq. (\ref{NTE})  are obtained 
by adding the data points of the cross-correlations between temperature and E-mode polarization \cite{mg1}.}
\begin{eqnarray}
&&(n_{\mathrm{B}},B_{\mathrm{L}})_{\Lambda\mathrm{CDM}} = ( 1.598,\,3.156 \mathrm{nG})\to (n_{\mathrm{B}} , B_{\mathrm{L}})_{\mathrm{{\it w}CDM}} =(1.883, \,4.982\, \mathrm{nG}), 
\label{NTT}\\
&&(n_{\mathrm{B}}, B_{\mathrm{L}})_{\Lambda\mathrm{CDM}} = ( 1.616,\,3.218 \mathrm{nG})\to (n_{\mathrm{B}},B_{\mathrm{L}})_{\mathrm{{\it w}CDM}} = (1.913, \,5.163\, \mathrm{nG}).
\label{NTE}
\end{eqnarray}
Even if the addition of a fluctuating dark energy background pins down systematically larger values of the magnetic field  parameters, the results of \cite{mg1,mg2} will be used here just for a consistent illustration
of the results. If  $n_{\mathrm{B}} > 1$ (as in Eqs. (\ref{NTT})--(\ref{NTE})) 
 a useful analytical approximation of the power spectra of $\sigma_{\mathrm{B}}(k,y)$ 
and $\Omega_{\mathrm{B}}(k,y)$ can be written as 
\begin{eqnarray}
&& \langle \Omega_{\rm B}(\vec{k},y) \Omega_{\rm B}(\vec{p},y)\rangle = \frac{2\pi^2 }{k^3} 
{\mathcal   P}_{\Omega}(k,y) \delta^{(3)}(\vec{k} + \vec{p}),
\nonumber\\
&& \langle \sigma_{\rm B} (\vec{k},y) \sigma_{\rm B}(\vec{p},y) \rangle = 
\frac{2 \pi^2}{k^3} {\mathcal   P}_{\sigma}(k,y) \delta^{(3)}(\vec{k} +\vec{p}),
\label{autoc}
\end{eqnarray}
where $\overline{\Omega}_{\mathrm{B\, L}} = B_{\mathrm{L}}^2/(8 \pi \overline{\rho}_{\gamma})$ and so the power 
spectra are given by:
\begin{equation}
{\mathcal   P}_{\Omega}(k,y) = {\mathcal   F}(n_{\mathrm{B}}, k_{\mathrm{D}}) \overline{\Omega}_{{\rm B\, L}}^2 \biggl(\frac{k}{k_{L}}\biggr)^{2 (n_{\mathrm{B}} -1)},
\qquad
{\mathcal   P}_{\sigma}(k,y) = {\mathcal   G}(n_{\mathrm{B}}, k_{\mathrm{D}}) \overline{\Omega}_{{\rm B\, L}}^2 \biggl(\frac{k}{k_{L}}\biggr)^{2 (n_{\mathrm{B}} -1)}.
\label{POM1}
\end{equation}
The functions $ {\mathcal   F}(n_{\mathrm{B}}, k_{\mathrm{D}}) $ and $ {\mathcal   G}(n_{\mathrm{B}}, k_{\mathrm{D}}) $ are defined in terms of the spectral index $n_{\mathrm{B}}$ and of the diffusive 
wavenumber $k_{\mathrm{D}}$:
\begin{eqnarray}
{\mathcal F}(n_{\mathrm{B}}, k_{\mathrm{D}}) &=& \frac{ 4 (7 - n_{\mathrm{B}})}{3 ( n_{\mathrm{B}} -1) ( 5 - 2 n_{\mathrm{B}})} - \frac{8 }{ 3 ( n_{\mathrm{B}} -1)} \biggl(\frac{k}{k_{\mathrm{p}}}\biggr)^{1 - n_{\mathrm{B}}} + 
\frac{4 }{ 2 n_{\mathrm{B}}-5} \biggl(\frac{k}{k_{\mathrm{D}}}\biggr)^{ 5 - 2 n_{\mathrm{B}}},
\nonumber\\
{\mathcal G}(n_{\mathrm{B}}, k_{\mathrm{D}}) &=& \frac{n_{\mathrm{B}} + 29}{15 ( 5 - 2 n_{\mathrm{B}}) ( n_{\mathrm{B}} -1)} - \frac{2}{3(n_{\mathrm{B}} -1)} \biggl(\frac{k}{k_{\mathrm{p}}}\biggr)^{1 - n_{\mathrm{B}}}
+ \frac{7}{5(2 n_{\mathrm{B}} - 5)}  \biggl(\frac{k}{k_{\mathrm{D}}}\biggr)^{ 5 - 2 n_{\mathrm{B}}}.
\label{POM2}
\end{eqnarray}
Recalling that the angular diameter distance at last scattering can be written as 
$D_{A}(z_{*}) = 2 d_{\mathrm{A}}(z_{*})/(H_{0} \sqrt{\Omega_{\mathrm{M}0}})$, $k_{\mathrm{A}}(z_{*})$ and 
$k_{\mathrm{D}}(z_{*})$ are determined in terms of the parameters of the fiducial set of parameters (given, for instance, by Eqs. (\ref{FL10}) and (\ref{FL12})):
\begin{equation}
k_{\mathrm{A}}(z_{*}) = 1/D_{\mathrm{A}}(z_{*}),\qquad \frac{k_{\mathrm{D}}(z_{*})}{k_{\mathrm{A}}(z_{*})}= \frac{2240 \, d_{\mathrm{A}}(z_{*})}{\sqrt{\sqrt{r_{\mathrm{R}*} +1} - \sqrt{r_{\mathrm{R}*}}}} 
\biggl(\frac{z_{*}}{10^{3}} \biggr)^{5/4} \, \omega_{\mathrm{b}0}^{0.24} \omega_{\mathrm{M}}^{-0.11}.
\label{PS4a}
\end{equation}
The results of Eqs. (\ref{POM1}) and (\ref{POM2}) compare pretty well with the numerical solution of the two convolutions defining 
${\mathcal   P}_{\Omega}(k,y)$ and ${\mathcal   P}_{\sigma}(k,y)$ \cite{mg4,mg6}. Note that, incidentally, 
the same kind of integrals determining ${\mathcal   P}_{\Omega}(k,y)$ and ${\mathcal   P}_{\sigma}(k,y)$ arise when computing 
secondary graviton spectra from waterfall fields where the recent numerical results of \cite{mg7} show, once more, 
 excellent agreement with the semi-analytical scheme leading to the results (\ref{POM1}) and (\ref{POM2}). 
The relative magnitude of $\Omega_{\mathrm{B\, L}}$ and of ${\mathcal A}_{{\mathcal R}}$ (i.e. 
the amplitude of the power spectrum of curvature perturbations) depends on $B_{\mathrm{L}}$
(i.e. the typical magnetic field intensity regularized over a typical length-scale $k_{\mathrm{L}}^{-1}$)
\begin{equation} 
\frac{\overline{\Omega}_{\mathrm{B L}}}{{\mathcal A}_{\mathcal R}}=  39.56\, \biggl(\frac{B_{\mathrm{L}}}{\mathrm{nG}}\biggr)^{2} \, \biggl(\frac{T_{\gamma0}}{2.725\,\mathrm{K}}\biggr)^{-4} \, \biggl(\frac{{\mathcal A}_{{\mathcal R}}}{2.41\times 10^{-9}}\biggr)^{-1}.
\label{PS6}
\end{equation}
In Fourier space and with all the specifications given above Eq. (\ref{GR10}) can be written as: 
\begin{equation}
\frac{d^2 \delta_{\mathrm{m}}}{d y^2} + \biggl[ \frac{1}{2} - \frac{3}{2} w_{\mathrm{de}} ( 1 - \Omega_{\mathrm{M}})\biggr] \frac{d \delta_{\mathrm{m}}}{d y} - \frac{3}{2} \Omega_{\mathrm{M}} \delta_{\mathrm{m}} = {\mathcal S}(k, y),
\label{GR10a}
\end{equation}
where 
\begin{equation}
{\mathcal S}(k,y) =  
\frac{k^2 \, \omega_{\mathrm{b}0}}{ R_{\mathrm{b}}(y) \,{\mathcal H}^2(y)\,\omega_{\mathrm{M}0} }\biggl[  \sigma_{\mathrm{B}}(k, y) 
-\frac{\Omega_{\mathrm{B}}(k,y)}{4}\biggr].
\label{GR10b}
\end{equation}
By introducing the explicit definition of the growth rate Eq. (\ref{GR10a}) and (\ref{GR10b}) 
lead to an integro-differential equation whose explicit form can be written as:
\begin{equation}
\frac{d f}{d y} + f^2 + \biggl[ \frac{1}{2} - \frac{3}{2} w_{\mathrm{de}} ( 1 - \Omega_{\mathrm{M}})\biggr] f - \frac{3}{2} \Omega_{\mathrm{M}} = 
 \frac{W_{\mathrm{B}}}{1 - 2 \,W_{\mathrm{B}}/3} \, \Omega_{\mathrm{M}}\, {\mathcal U}[f;\, y_{\mathrm{i}},\, y],
\label{GR11}
\end{equation}
where ${\mathcal U}[f;\, y_{\mathrm{i}},\, y]$ is a functional of the growth rate; $W_{\mathrm{B}} =
W_{\mathrm{B}}(k, n_{\mathrm{B}}, B_{\mathrm{L}})$ depends upon the magnetic field spectra and upon 
the spectrum of matter inhomogeneities at $y_{\mathrm{i}}$:
\begin{eqnarray}
&& {\mathcal U}[f;\, y_{\mathrm{i}},\, y] = e^{- I(y_{\mathrm{i}},\, y)}, \qquad 
I(y_{\mathrm{i}},\, y) = \int_{y_{\mathrm{i}}}^{y} \, f(x)\, dx,
\label{defU1}\\
&&W_{\mathrm{B}}(k, n_{\mathrm{B}}, B_{\mathrm{L}}) = \frac{\omega_{\gamma0}}{3 \omega_{\mathrm{M}0}^2 \, | \delta_{\mathrm{m}}(k,y_{\mathrm{i}})|} 
\, \biggl(\frac{k \, h_{0}}{H_{0}}\biggr)^{2} \biggl[ 4|\sigma_{\mathrm{B}}(k)| - |\Omega_{\mathrm{B}}(k)|\biggr].
\label{GR11W}
\end{eqnarray}
For typical scales $k \gg  k_{\mathrm{eq}}$ the matter power spectrum at $y_{\mathrm{eq}}$ can be approximated
 as\footnote{In Eq. (\ref{GR11b}) $\delta_{H}(k)$ denotes  the initial spectrum which is related to the spectrum of curvature  perturbations while $T(k/k_{\mathrm{eq}})$  denotes the transfer function.}
\begin{equation}
{\mathcal P}_{\delta}(k,y_{\mathrm{eq}}) = T^2(k/k_{\mathrm{eq}}) \delta_{\mathrm{H}}^2(k) \to \frac{4}{25}{\mathcal A}_{{\mathcal R}} \biggl(\frac{k}{k_{\mathrm{p}}}\biggr)^{n_{\mathrm{s}} -1} \ln^2{(k/k_{\mathrm{eq}})},
\label{GR11b}
\end{equation}
so that Eq. (\ref{GR11W}) can also be written as: 
\begin{eqnarray}
W_{\mathrm{B}}(k, n_{\mathrm{B}}, B_{\mathrm{L}}) &=& \frac{5\, \omega_{\gamma0}}{6\, \omega_{\mathrm{M}0}^2} 
\biggl(\frac{k}{k_{\mathrm{L}}}\biggr)^{n_{\mathrm{B}} -1}\, \biggl(\frac{k}{k_{\mathrm{p}}}\biggr)^{( 1 - n_{\mathrm{s}})/2}
\frac{\overline{\Omega}_{\mathrm{B}\mathrm{L}}}{\sqrt{{\mathcal A}_{\mathcal R}}} \, \frac{k^2 \, h_{0}^2}{H_{0}^2\, \ln{(k/k_{\mathrm{eq}})}}\, {\mathcal L}_{\mathrm{B}}(n_{\mathrm{B}}, k_{\mathrm{D}}),
\nonumber\\
 {\mathcal L}_{\mathrm{B}}(n_{\mathrm{B}}, k_{\mathrm{D}}) &=&
 \frac{(2 \pi)^{n_{\mathrm{B}} -1}}{\Gamma[(n_{\mathrm{B}} -1)/2]} 
 \biggl[ 4\sqrt{|{\mathcal G}(n_{\mathrm{B}}, k_{\mathrm{D}})| }- \sqrt{|{\mathcal F}(n_{\mathrm{B}}, k_{\mathrm{D}})|}\biggr].
\label{GR11c}
\end{eqnarray}
The evolution variable in Eq. (\ref{GR11}) is $y$ but since the relation between  $\Omega_{\mathrm{M}}$ and $y$
is regular and invertible, Eq. (\ref{FL9}) can be used to obtain the evolution of $f$ directly in terms of $\Omega_{\mathrm{M}}$:
\begin{eqnarray}
&&3 w_{\mathrm{de}} (1 - \Omega_{\mathrm{M}}) \frac{ d f}{d \ln{\Omega_{\mathrm{M}}}} + f^2 + 
\biggl[ \frac{1}{2} - \frac{3}{2} w_{\mathrm{de}} ( 1 - \Omega_{\mathrm{M}})\biggr] f - \frac{3}{2} \Omega_{\mathrm{M}}   = W_{\mathrm{B}}\,{\mathcal U}[f;\Omega_{\mathrm{M}}], 
\label{GR13}
\end{eqnarray}
where 
\begin{equation}
I(\Omega_{\mathrm{M}}) = \frac{1}{3 w_{\mathrm{de}}} \int_{\Omega_{\mathrm{M}}(y_{\mathrm{i}})}^{\Omega_{\mathrm{M}}} \frac{f(\Omega)}{\Omega(1 - \Omega)}
d\Omega, \qquad W_{\mathrm{B}} \equiv W_{\mathrm{B}}(k, n_{\mathrm{B}}, B_{\mathrm{L}}).
\label{GR13a}
\end{equation}
The solution 
of Eq. (\ref{GR11}) and (\ref{GR13}) will now be parametrized as
\begin{equation}
f(k,y) = \frac{\overline{f}(y)}{ 1 - 2 W_{\mathrm{B}} \, {\mathcal U}[\overline{f}\,; y_{\mathrm{i}},\, y]/3},
\label{Nw1}
\end{equation}
where $\overline{f}(y)$ obeys the following equation:
\begin{equation}
\frac{d \overline{f}}{d y} + \overline{f}^2 + \biggl[ \frac{1}{2} - \frac{3}{2} w_{\mathrm{de}} ( 1 - \Omega_{\mathrm{M}})\biggr] \overline{f} - \frac{3}{2} \Omega_{\mathrm{M}} =0,
\label{Nw2}
\end{equation}
and where ${\mathcal U}[\overline{f}\,; y_{\mathrm{i}},\, y]$ has been defined in Eq. (\ref{defU1}); notice, however, 
that in Eq. (\ref{Nw1}) ${\mathcal U}$ is a functional of $\overline{f}$ and not simply of $f$.
To prove that Eqs. (\ref{Nw1}) and (\ref{Nw2}) indeed solve Eq. (\ref{GR11}) and (\ref{GR13}) let us 
start by showing that if Eq. (\ref{Nw1}) holds, then the following relation is also verified:
\begin{equation}
{\mathcal U}[f\,; y_{\mathrm{i}},\, y] = \frac{(1 - 2 W_{\mathrm{B}}/3){\mathcal U}[\overline{f}\,; y_{\mathrm{i}},\, y] }{1 - 2 W_{\mathrm{B}} \, {\mathcal U}[\overline{f}\,; y_{\mathrm{i}},\, y]/3}.
\label{Nw3}
\end{equation}
Equation (\ref{Nw3}) follows by integrating Eq. (\ref{Nw1}) over a dummy variable (be it $x$) between
$y_{\mathrm{i}}$ and $y$; let us then show this explicitly and first rewrite Eq. (\ref{Nw1}) as 
\begin{equation}
f(k, x) = \overline{f}(x) + \frac{2 W_{\mathrm{B}} {\mathcal U}[\overline{f}\,; x_{\mathrm{i}},\, x] \overline{f}(x)/3}{1 
- 2  W_{\mathrm{B}} {\mathcal U}[\overline{f}\,; x_{\mathrm{i}},\, x]/3};
\label{Nw4}
\end{equation}
note that, at the right hand side of Eq. (\ref{Nw4}), the argument of ${\mathcal U}$ is $\overline{f}$ (and not 
$f$). By integrating over $x$ the left and right hand sides of Eq. (\ref{Nw4}) 
between $y_{i}$ and $y$ the following equation can be easily obtained:
\begin{equation}
- \int_{y_{\mathrm{i}}}^{y} f(k, x)\, d x = -  \int_{y_{\mathrm{i}}}^{y} \overline{f}( x)\, d x - 
\ln{\biggl[ \frac{1 - 2  W_{\mathrm{B}} {\mathcal U}[\overline{f}\,; y_{\mathrm{i}},\, y]/3}{1 - 2  \,W_{\mathrm{B}}/3}\biggr]},
\label{Nw5}
\end{equation}
where the logarithmic contribution arises by direct integration of the second term in Eq. (\ref{Nw4}) between the two limits 
$y_{\mathrm{i}}$ and $y$.
The left and the right hand sides of Eq. (\ref{Nw5}) can then be exponentiated and 
Eq. (\ref{Nw3}) is recovered. Note that, according to Eq. (\ref{Nw3}), ${\mathcal U}[f\,; y_{\mathrm{i}},\, y_{\mathrm{i}}] =1$ 
as it must be for consistency with the definition of ${\mathcal U}[f\,; y_{\mathrm{i}},\, y]$ given in Eq. (\ref{defU1}).
Bearing now in mind the results of Eq. (\ref{Nw3}), 
$f(k,y)$ given by Eq. (\ref{Nw1}) can be plugged into Eq. (\ref{GR11}) and the obtained equation 
for $\overline{f}(y)$ turns out to be exactly the one given in Eq. (\ref{Nw2}). 

\begin{figure}[!ht]
\centering
\includegraphics[height=6cm]{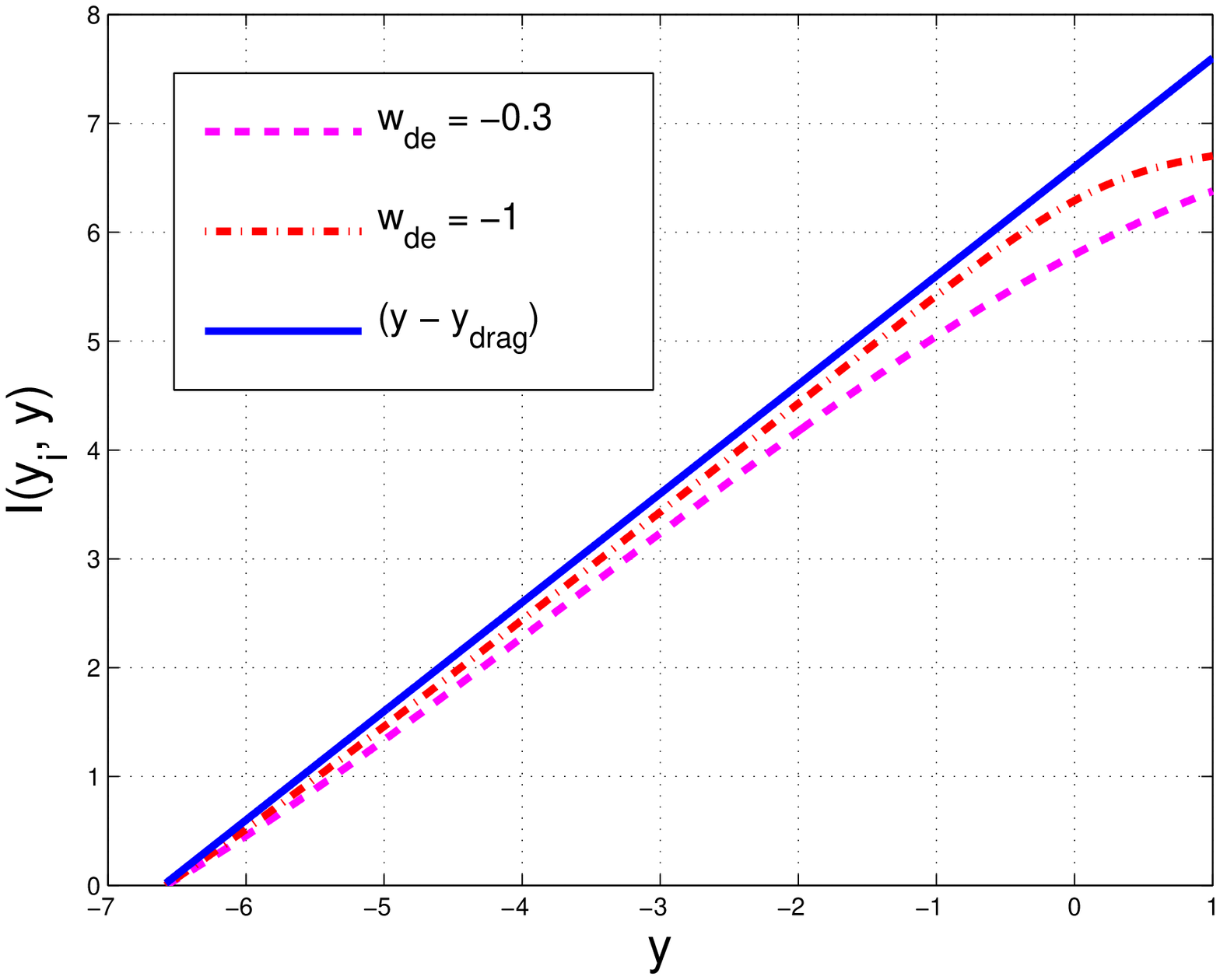}
\includegraphics[height=6cm]{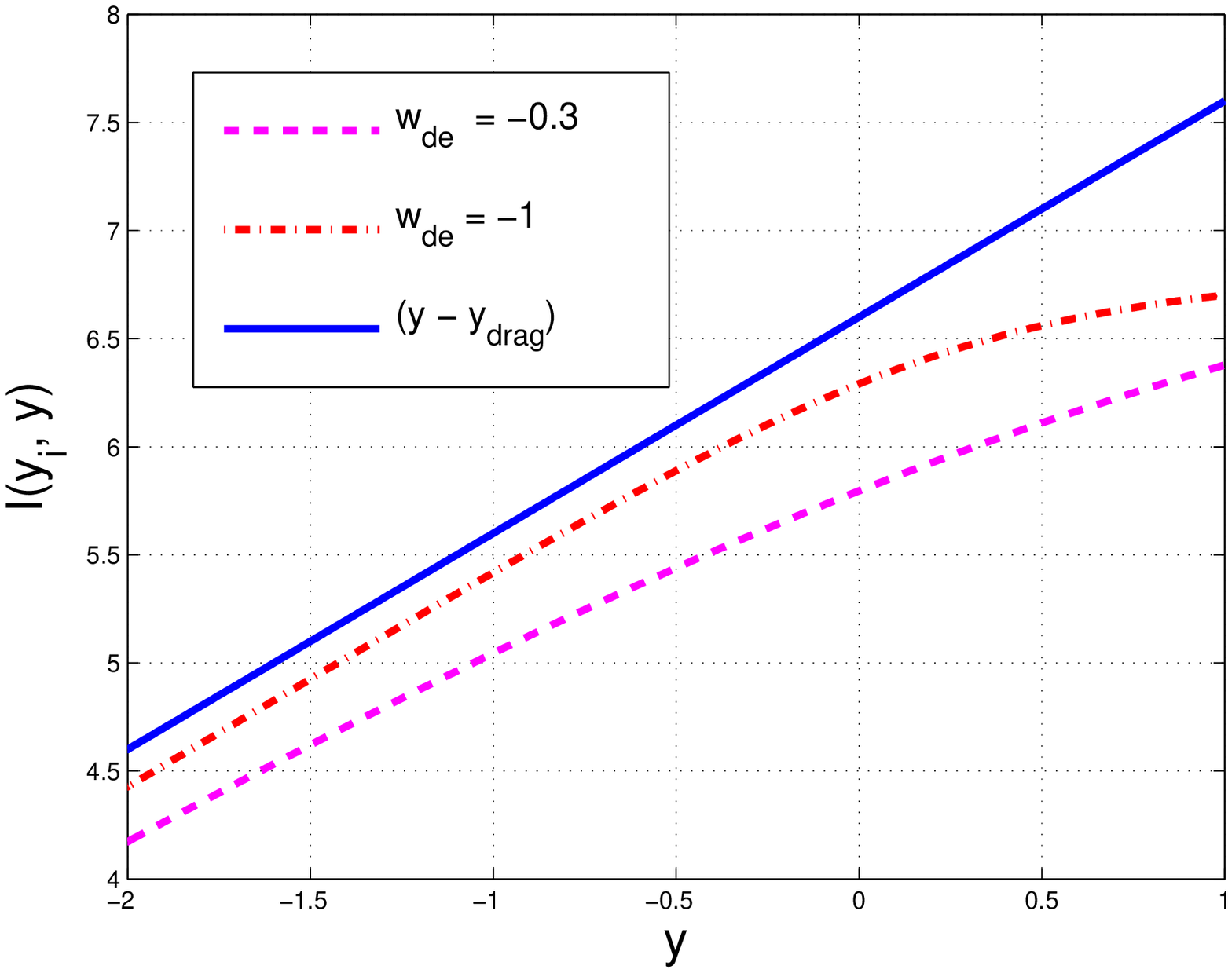}
\caption[a]{The result of the numerical evaluation of the integral of Eq. (\ref{GR15b}) is compared with the analytical interpolation discussed in the text (plot at the left).
The agreement is reasonable for most of the ranges of $w_{\mathrm{de}}$ and $y$ but it gets worse (see the right 
plot) in the region of large $y$ (i.e. $y \to 1$) and large $w_{\mathrm{de}}$ (i.e. $w_{\mathrm{de}} \to - 0.3$).}
\label{figure1}      
\end{figure}

To derive an explicit solution for the growth rate we need to solve Eq. (\ref{Nw2}), but this part of the problem is more conventional. By assuming that $\overline{f} = \Omega_{\mathrm{M}}^{\gamma}$, Eq. (\ref{Nw2}) becomes:
\begin{equation}
3 w_{\mathrm{de}} ( 1 - \Omega_{\mathrm{M}})\biggl[ \ln{\Omega_{\mathrm{M}}} \frac{d \gamma}{d \ln{\Omega_{\mathrm{M}}}}  + \gamma\biggr] \Omega_{\mathrm{M}}^{\gamma}
+ \Omega_{\mathrm{M}}^{2\gamma} + \biggl[ \frac{1}{2} - \frac{3}{2} w_{\mathrm{de}} ( 1 - \Omega_{\mathrm{M}})\biggr] \Omega_{\mathrm{M}}^{\gamma} - \frac{3}{2}\Omega_{\mathrm{M}}=0.
\label{GR14}
\end{equation}
Eq. (\ref{GR14}) can be solved as a power series in $\epsilon = 1 - \Omega_{\mathrm{M}}$
 by neglecting, to lowest order in $\epsilon$, the derivatives of $\gamma$ with respect to $\Omega_{\mathrm{M}}$.  A posteriori this assumption will 
turn out to be rather accurate and it is commonly adopted in the absence of magnetic field contribution \cite{gamma2,gamma3}. Thus, from Eq. (\ref{GR14}) the following relation can be determined:
\begin{equation}
3 w_{\mathrm{de}} ( 1 - \Omega_{\mathrm{M}}) \gamma + \Omega_{\mathrm{M}}^{\gamma}+ 
 \biggl[ \frac{1}{2} - \frac{3}{2} w_{\mathrm{de}} ( 1 - \Omega_{\mathrm{M}})\biggr] - \frac{3}{2}\Omega_{\mathrm{M}}^{1-\gamma}=0.
\label{GRR14}
\end{equation}
By expanding the left hand side we obtain, to leading order in $\epsilon = 1 - \Omega_{\mathrm{M}}$, 
\begin{equation}
\gamma = \frac{ 3 w_{\mathrm{de}} - 3}{6 w_{\mathrm{de}} - 5} 
+ \frac{3}{125} \frac{(1- w_{\mathrm{de}})(1 - 3 w_{\mathrm{de}}/2)}{(1- 6 w_{\mathrm{de}}/5)^2}\epsilon +  {\mathcal O}(\epsilon^2).
\label{GR15}
\end{equation}
Inserting the leading order result of Eq. (\ref{GR15}) into Eq. (\ref{Nw1}) the explicit form of the growth rate becomes 
\begin{equation}
f(k,y) = \frac{\Omega_{\mathrm{M}}^{\gamma}}{ 1 - 2 W_{\mathrm{B}} \, 
{\mathcal U}[\overline{f}\,; y_{\mathrm{i}},\, y]/3},\qquad {\mathcal U}[\overline{f}\,; y_{\mathrm{i}},\, y] = 
\exp{\biggl[ - \int_{y_{\mathrm{i}}}^{y} \Omega_{\mathrm{M}}^{\gamma} (x)dx\biggr]},
\label{GR15a}
\end{equation}
where, now,  $\gamma= (3 w_{\mathrm{de}} - 3)/(6 w_{\mathrm{de}} - 5) + {\mathcal O}(\epsilon)$ and, as already mentioned 
prior to Eq. (\ref{GR11W}),  $W_{\mathrm{B}}= W_{\mathrm{B}}(k, n_{\mathrm{B}}, B_{\mathrm{L}})$ is given by Eq. (\ref{GR11c}). Even if more accurate analytical 
results can be obtained by including higher orders in $\epsilon$, to privilege the simplicity 
of the approach and of the comparison we will stick, for the present treatment, to the explicit expression of Eq. (\ref{GR15a}) with 
the growth index evaluated to lowest order in $\epsilon$. 
The integral appearing in Eq. (\ref{GR15a})  can be performed either directly 
or by using $\Omega$ as integration variable (see, e.g. Eq. (\ref{FL9}) for the appropriate change of integration variables):
\begin{equation}
\int_{y_{\mathrm{i}}}^{y} \Omega_{\mathrm{M}}^{\gamma}(x) \, dx = \frac{1}{3 w_{\mathrm{de}}} 
\int_{\Omega_{\mathrm{M}}(y_{\mathrm{i}})}^{\Omega_{\mathrm{M}}} \frac{\Omega^{\gamma}}{\Omega \, (\Omega -1)} \, d\Omega. 
\label{GR15b}
\end{equation}
The lower limit of integration indicated in Eq. (\ref{GR14}) can be estimated by recalling that
\begin{equation}
\Omega_{\mathrm{M}}(y_{\mathrm{eq}}) = \biggl[ 2 + \frac{\Omega_{\mathrm{R}0}}{\Omega_{\mathrm{M}0}} 
\biggl(\frac{\Omega_{\mathrm{de}0}}{\Omega_{\mathrm{M}0}}\biggr)^{\frac{1}{3 w_{\mathrm{de}}}}\biggr]^{-1} \to \frac{1}{2},\qquad \frac{\Omega_{\mathrm{R}0}}{\Omega_{\mathrm{M}0}} \simeq {\mathcal O}(10^{-5}).
\end{equation}
Thus, if $ y_{\mathrm{eq}} \ll y_{\mathrm{i}} \ll y_{\mathrm{de}}$, $\Omega_{\mathrm{M}}(y_{\mathrm{i}}) > 1/2$.
 For instance 
$y_{\mathrm{eq}} = - 7.78$ for the fiducial set of parameters of Eq. (\ref{FL10}). Thus $y_{\mathrm{i}}$ 
can be safely chosen in the range $ -7 < y_{\mathrm{i}} < -4$ if we want to be consistent with the 
approximations made so far.
In Fig. \ref{figure1} the result of the numerical evaluation of Eq. (\ref{GR15b}) is reported in the two physical cases which have been 
taken as extreme, i.e. $w_{\mathrm{de}} = -1$ and $w_{\mathrm{de}} =-0.3$. With the continuous line the analytical interpolation 
$I(y_{\mathrm{i}}, y) = (y - y_{\mathrm{drag}})$ is reported.  
The fiducial set of the cosmological parameters coincides with 
the one reported in Eq. (\ref{FL10}). The values of $B_{\mathrm{L}}$ and $n_{\mathrm{B}}$ 
 lie within the range of parameters determined from the analysis temperature and polarization anisotropies of the CMB within the magnetized $\Lambda$CDM scenario and in the light of the WMAP data alone \cite{mg1,mg2} (see also Eqs. (\ref{NTT}) and (\ref{NTE})).

\begin{figure}[!ht]
\centering
\includegraphics[height=6cm]{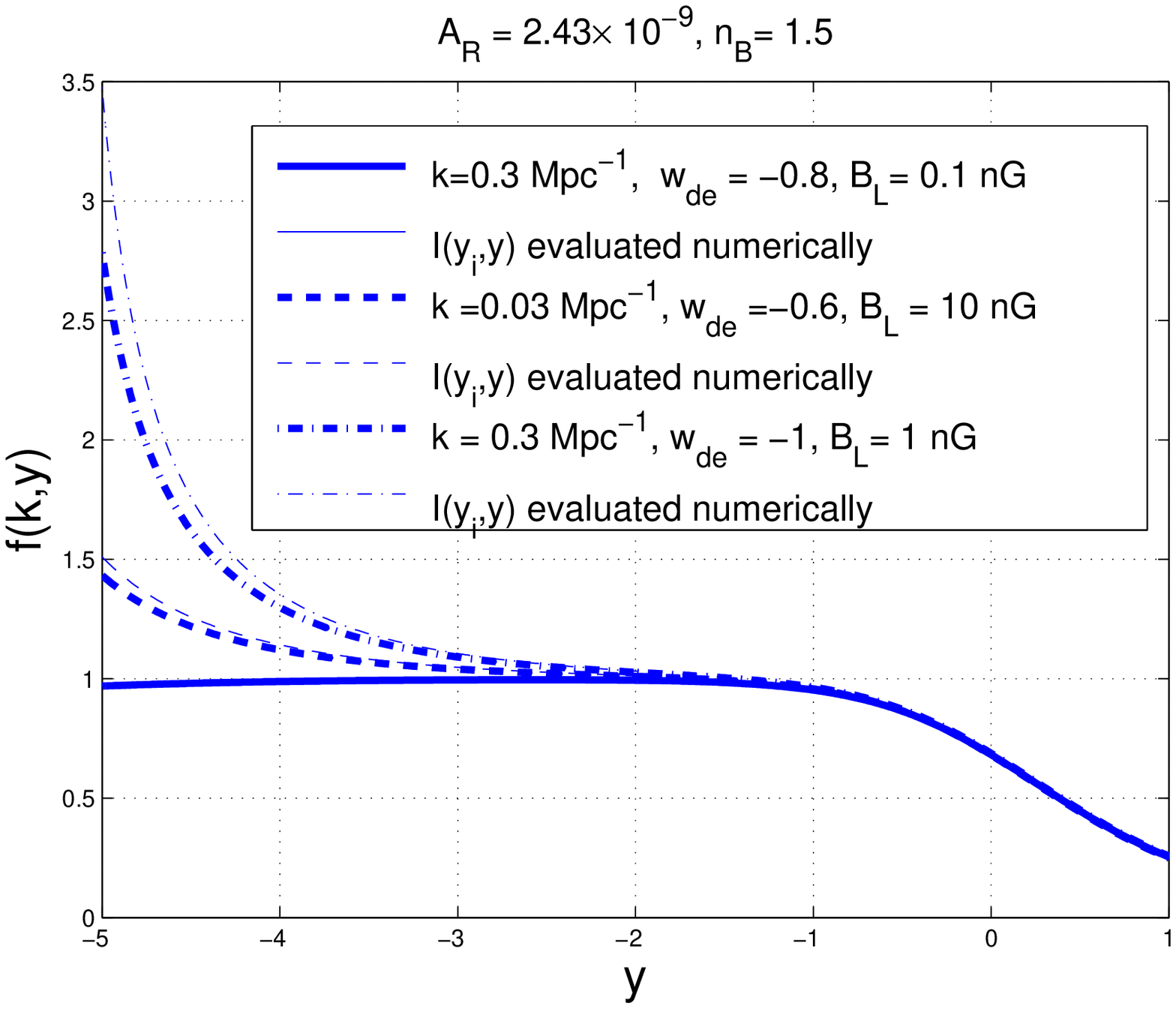}
\includegraphics[height=6cm]{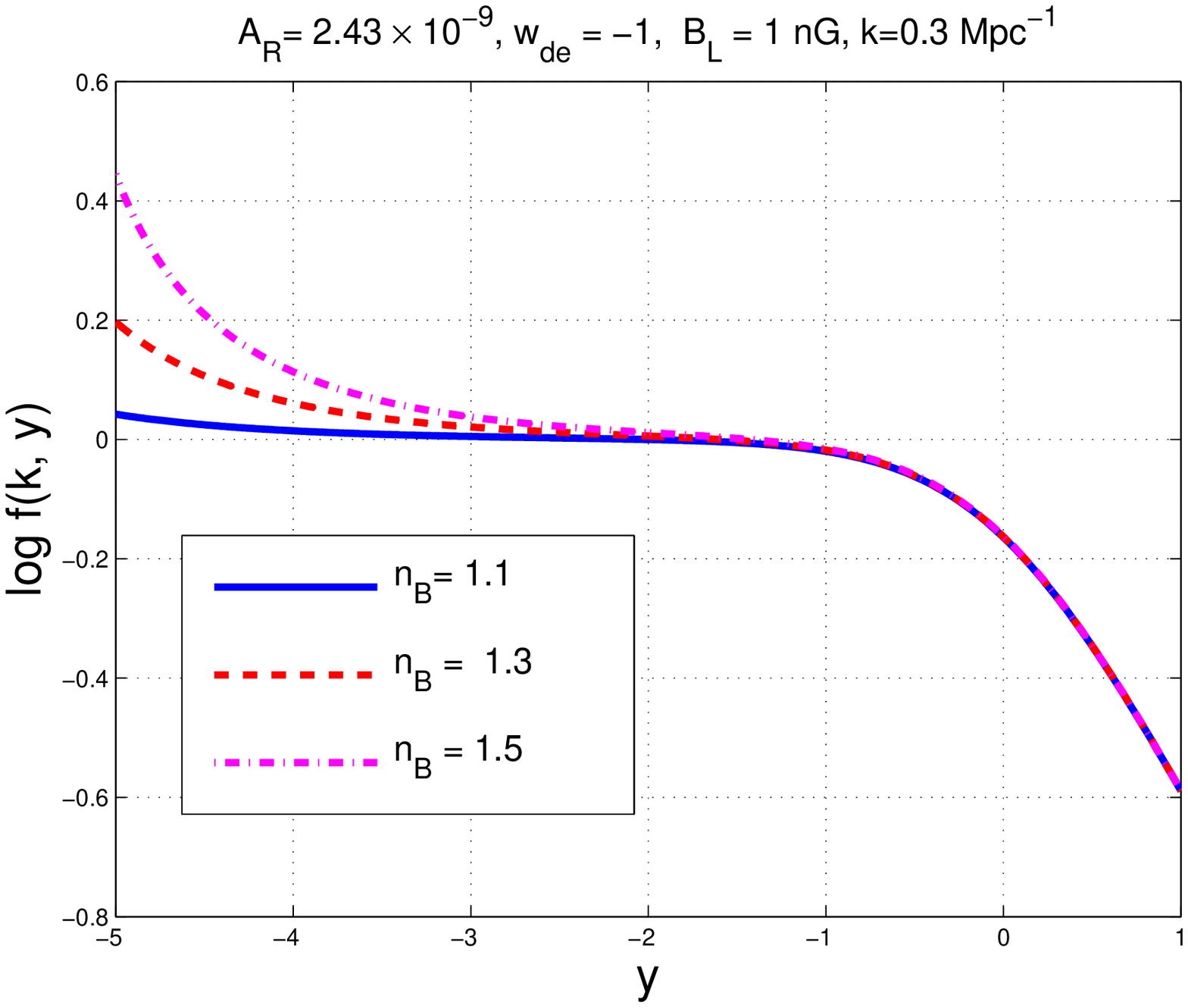}
\caption[a]{The growth factor is illustrated analytically and numerically (plot at the left) for different values of the comoving wavenumber and of the magnetic field intensity but 
keeping the magnetic spectral index $n_{\mathrm{B}}$ fixed. In the right plot the common logarithm of the growth rate is illustrated for different values of the magnetic spectral indices by fixing all the other parameters, as explained in the text.}
\label{figure2}      
\end{figure}

\begin{figure}[!ht]
\centering
\includegraphics[height=6.5cm]{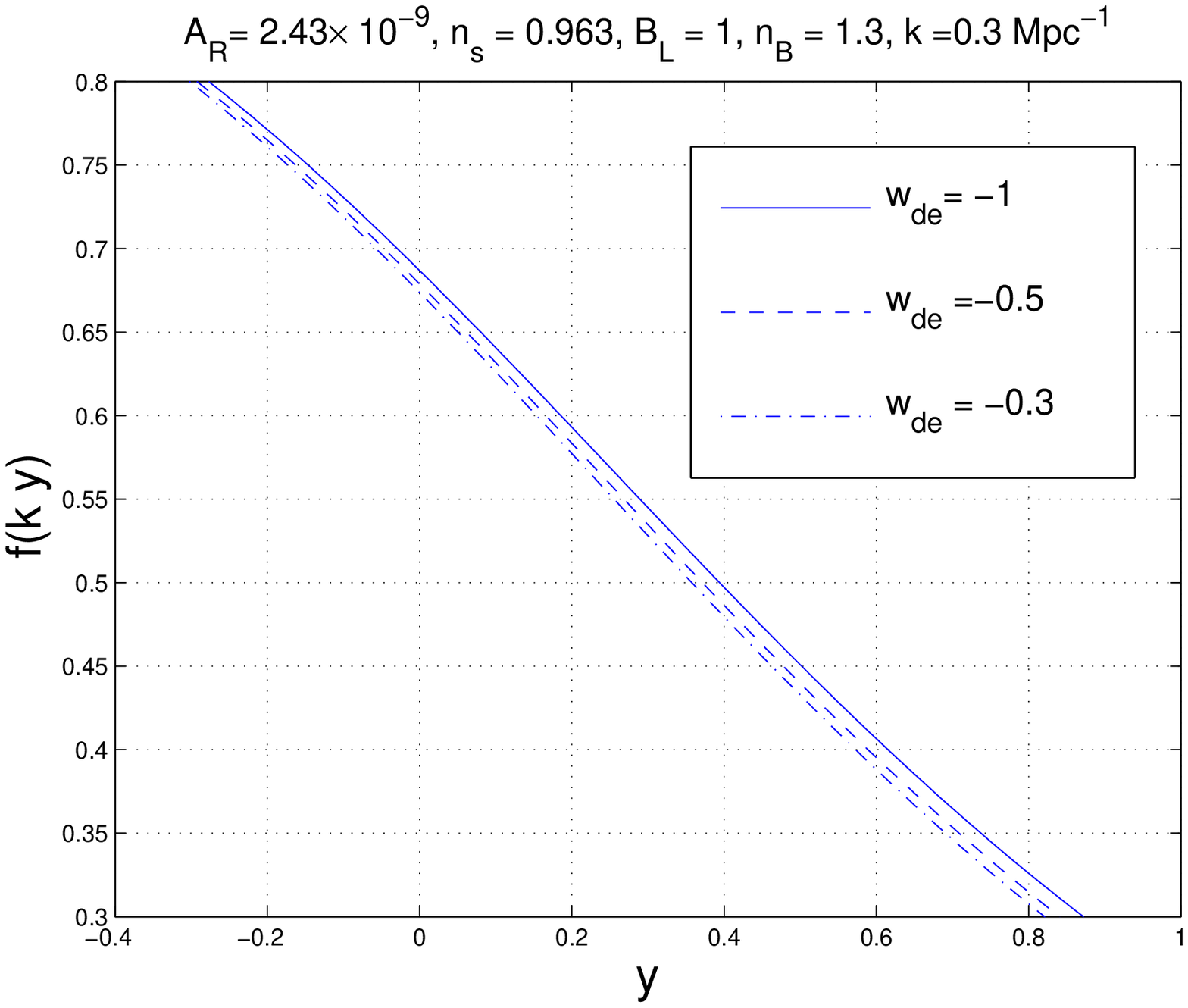}
\includegraphics[height=6.5cm]{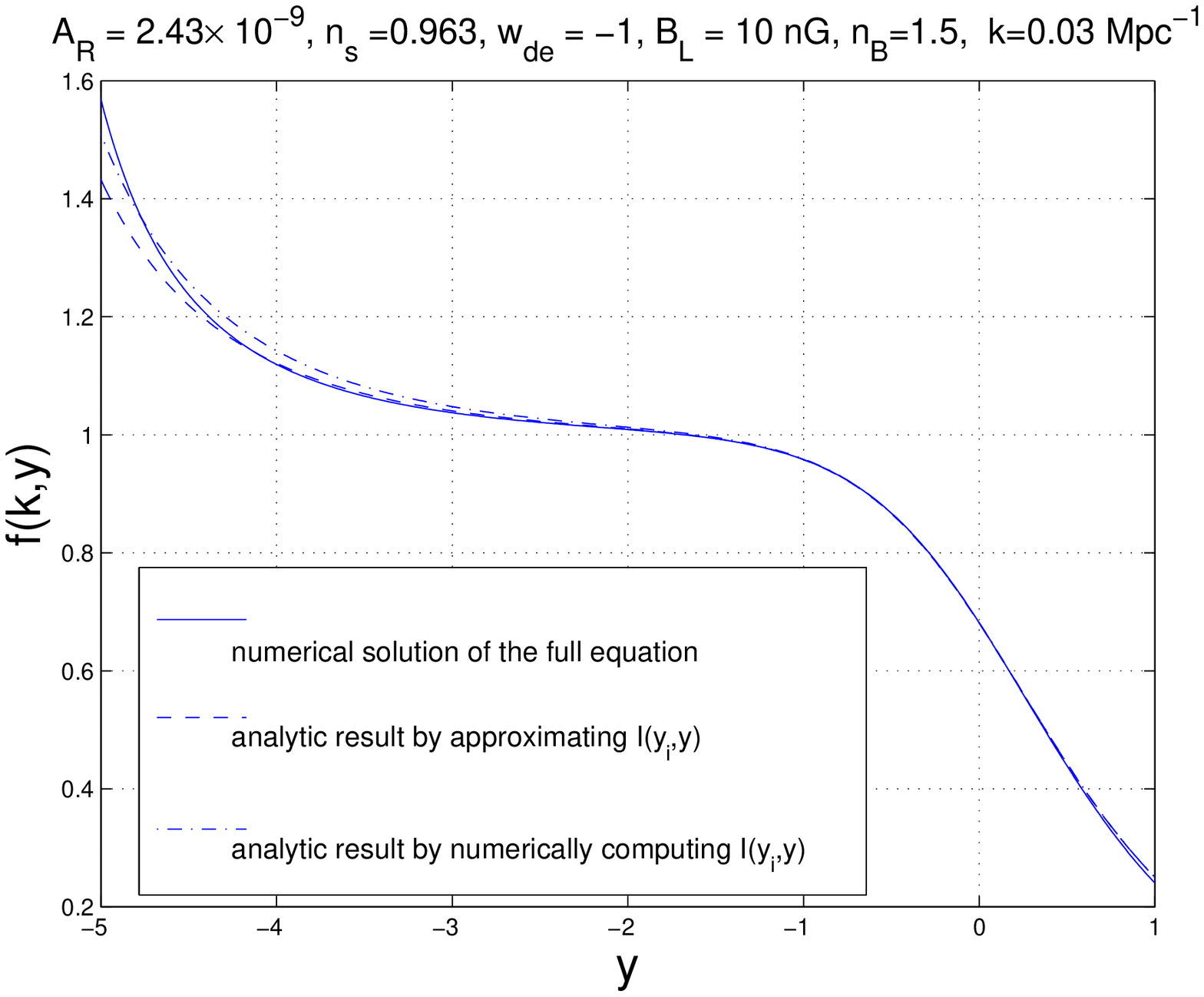}
\caption[a]{The growth factor for different values of $w_{\mathrm{de}}$ (plot at the left);  the full numerical solution 
of the evolution equation of the density contrast is compared with the analytic and semi-analytic approximations previously discussed (plot at the right).}
\label{figure3}      
\end{figure}

In  Fig. \ref{figure2} (plot at the left) the growth factor of Eqs. (\ref{Nw1}) and (\ref{GR15a})--(\ref{GR15b}) is illustrated.  The thick lines in Fig. \ref{figure2} correspond to the numerical evaluation of the integral of Eq. (\ref{GR15b}) while the thin lines are obtained by means of the analytic approximation (full line in Fig. \ref{figure1}). The analytic approximation is adequate for the purposes of the present analysis. In the right plot in Fig. \ref{figure2} the common logarithm of the growth factor is illustrated for a fixed scale and fixed magnetic field amplitude but different spectral indices. 
In the left plot of Fig. \ref{figure3} the growth rate Eqs. (\ref{Nw1}) and (\ref{GR15a})--(\ref{GR15b}) is plotted 
for different values of $w_{\mathrm{de}}$ and by fixing all the other parameters to a selected fiducial value mentioned 
in the legends and in the labels appearing at the top of each plot.   The range of $y$ has been narrowed in the plot so that the curves corresponding to different values of $w_{\mathrm{de}}$ are more clear.
In the right plot of Fig. \ref{figure3} the direct numerical solution of Eq. (\ref{GR10}) 
is illustrated with a full (thin) line. The dashed and dot-dashed lines denote, respectively, the semi-analytic and the fully analytic result. By semi-analytic result we mean the expression for the growth rate 
obtained by estimating numerically the integral of Eqs. (\ref{Nw1}) and (\ref{GR15b}) (see also dot-dashed 
and dashed lines in Fig.\ref{figure1}). In the analytic case the integral $I(y,y_{\mathrm{i}})$ is 
estimated by the approximation illustrated, with the full line, in Fig.\ref{figure1}. The direct numerical solution of Eq. (\ref{GR10}) leads to a result which is approximately located between the analytical and semi-analytical approximations. Other 
choices of the relevant parameters lead to results whose accuracy is comparable with the encouraging 
results of Fig.\ref{figure2}.
The only caveat with the latter statement is that, of course,  Eqs. (\ref{Nw1}) and (\ref{GR15a})--(\ref{GR15b}) 
are in good agreement with the numerical solution of the {\em approximate} equation (i.e. Eq. (\ref{GR10})).
This aspect will be further deepened in section \ref{sec4}.

The results reported so far have been derived within the synchronous gauge description. So, for instance, in a different 
gauge Eq. (\ref{GR7}) will assume a different form since $\delta_{\mathrm{m}}$ is not gauge invariant. This is 
not a problem since once the calculation is performed in a given gauge, the results can be translated in a different coordinate system. 
Still it is relevant to point out  that the growth equation 
obtained in Eq. (\ref{GR7}) is not gauge-invariant since the matter density contrast $\delta_{\mathrm{m}}$
does change for infinitesimal coordinate transformations. Very often, indeed, Eq. (\ref{GR7}) is derived in 
the conformally Newtonian gauge but this demands, as we shall show, different assumptions 
on the relative smallness of the relativistic corrections. 
So, even if this is a common problem also in the 
conventional case the considerations developed hereunder seem to be appropriate. 
The same steps leading to Eq. (\ref{GR7}) can be repeated in the longitudinal gauge with the result\footnote{Note that $\overline{\delta}_{X}$ denotes the density contrast of the species $X$ in the longitudinal 
gauge.}:
\begin{equation} 
\overline{\delta}_{\mathrm{m}}'' + {\mathcal H} \overline{\delta}_{\mathrm{m}}' = 3 (\psi'' + {\mathcal H} \psi') 
+\nabla^2 \phi -  \frac{\omega_{\mathrm{b}0}}{\omega_{\mathrm{M}0}} \frac{\vec{\nabla} \cdot 
( \vec{J} \times \vec{B})}{a^4 \, \rho_{\mathrm{b}}},
\label{GR16}
\end{equation}
where $\overline{\delta}_{\mathrm{m}}$ now denotes the density contrast in the longitudinal gauge; the definition 
of $\phi$ and $\psi$ can be found in Eq. (\ref{gauge2}). Using the Hamiltonian constraint in the longitudinal gauge, i.e. 
\begin{equation}
\nabla^2 \psi = 3 {\mathcal H} (\psi' + {\mathcal H} \phi) + \frac{3}{2} {\mathcal H}^2 \biggl[ 
 \Omega_{\mathrm{M}} \overline{\delta}_{\mathrm{m}} + \Omega_{\mathrm{R}} \overline{\delta}_{\mathrm{R}} +
\Omega_{\mathrm{de}} \overline{\delta}_{\mathrm{de}} + R_{\gamma} \Omega_{\mathrm{R}} \Omega_{\mathrm{B}} \biggr],
\label{GR17}
\end{equation}
as well as the dynamical equation for $\psi$, Eq. (\ref{GR16}) becomes
\begin{eqnarray}
&&\overline{\delta}_{\mathrm{m}}'' + {\mathcal H} \overline{\delta}_{\mathrm{m}}' = 
- 3 ({\mathcal H} \psi' + 2 {\mathcal H}' \psi) -  \frac{\omega_{\mathrm{b}0}}{\omega_{\mathrm{M}0}} \frac{\vec{\nabla} \cdot ( \vec{J} \times \vec{B})}{a^4 \, \rho_{\mathrm{b}}}
\nonumber\\
&& + \frac{3}{2} {\mathcal H}^2 \biggl[ 2 \Omega_{\mathrm{R}} \overline{\delta}_{\mathrm{R}} 
+ \Omega_{\mathrm{M}} \overline{\delta}_{\mathrm{m}} 
+ 2 R_{\gamma} \Omega_{\mathrm{R}} \Omega_{\mathrm{B}} + \Omega_{\mathrm{de}}( 1 + 3 c_{\mathrm{de}}^2) - \frac{9\, \Omega_{\mathrm{de}}}{{\mathcal K}(y)}
w_{\mathrm{de}} 
(w_{\mathrm{de}} + 1) \overline{\theta}_{\mathrm{de}}\biggr],
\label{GR18}
\end{eqnarray}
which coincides with Eq. (\ref{GR7}) only in the non-relativistic limit (i.e. $|\nabla^2 \psi| \gg \psi''$ and 
$|\nabla^2 \psi| \gg {\mathcal H} \psi'$) and in the case $\psi \to \phi$.  It could be naively expected that 
the same physical approximations leading to the growth equation in one gauge would lead to the same
growth equation in another gauge. This naive expectation is incorrect 
as Eqs. (\ref{GR7}) and (\ref{GR18}) show. The correct conclusion drawn from the comparison 
of Eqs. (\ref{GR7}) and (\ref{GR18}) is that the non-relativistic limit is implemented 
in different ways in different coordinate systems. The equation for $\delta_{\mathrm{m}}$, valid in the 
synchronous gauge, holds under milder assumptions, i.e. $\Omega_{\mathrm{R}} \ll \Omega_{\mathrm{M}}$
and  $\delta_{\mathrm{de}} \simeq 0$. The equation for $\overline{\delta}_{\mathrm{m}}$, under the same assumptions, also contains extra terms which can only be neglected 
by enforcing the non-relativistic limit in Eq. (\ref{GR18}) and by consequently neglecting $3 ({\mathcal H} \psi' + 2 {\mathcal H}' \psi) $. The synchronous description seems therefore 
more appropriate for the consistent computation of the growth rate. 
It might seem that the problem of the ambiguity in the definition of the growth equation 
could be solved by appealing to the standard gauge-invariant descriptions. 
 Consider for instance a conventional set of gauge-invariant variables such as the 
standard gauge-invariant generalization of the longitudinal gauge variables i.e.
$\psi \to \Psi$, $\phi \to \Phi$ and $\overline{\delta}_{X}\to \delta_{X}^{(\mathrm{gi})}$ where 
$\Phi$ and $\Psi$ are the two Bardeen potentials. In this case the equation for the gauge-invariant 
density contrast $\delta_{\mathrm{m}}^{(\mathrm{gi})}$ will be the same as Eq. (\ref{GR18}).
It seems more interesting to introduce the gauge-invariant variables (see e.g. \cite{mg6}, first paper)
\begin{equation}
\zeta_{\mathrm{c}} = - \psi + \frac{\overline{\delta}_{\mathrm{c}}}{3}, \qquad 
\zeta_{\mathrm{b}} = - \psi + \frac{\overline{\delta}_{\mathrm{b}}}{3},
\label{GR19}
\end{equation}
which are essentially the density contrasts of the the CDM and of the baryons but on the 
hypersurface where the curvature in unperturbed (see, e.g. \cite{hwang1,hwang2}).
Recalling, from the Hamiltonian constraint, that $\nabla^2 \psi = 12 \pi G a^2 (\rho_{\mathrm{t}} + p_{\mathrm{t}}) ( \zeta - {\mathcal R})$ we have that the evolution equation for 
$\zeta_{\mathrm{m}}= (\omega_{\mathrm{c}0}/\omega_{\mathrm{M}0}) \zeta_{\mathrm{c}} +  
(\omega_{\mathrm{b}0}/\omega_{\mathrm{M}0}) \zeta_{\mathrm{b}}$ becomes 
\begin{equation}
\zeta_{\mathrm{m}}'' + {\mathcal H} \zeta_{\mathrm{m}}' = ({\mathcal H}^2 - {\mathcal H}') 
( \zeta - {\mathcal R} ) - 3 \frac{\omega_{\mathrm{b}0}}{\omega_{\mathrm{M}0}} \frac{\vec{\nabla}
 \cdot (\vec{J} \times \vec{B})}{a^4 \rho_{\mathrm{b}}},\qquad \zeta = \sum_{\mathrm{a}} \frac{\rho_{\mathrm{a}}'}{\rho_{\mathrm{t}'}} \zeta_{\mathrm{a}},
 \label{GR20}
\end{equation}
where ${\mathcal R}$ is the standard variable describing the curvature perturbations on in the 
comoving orthogonal gauge and $\zeta$ is  the total density contrast in the uniform 
curvature gauge. This description can be used for the computation of the growth rate with 
some advantages which are, however, not central to the present discussion. 

Before concluding this section it is useful to is useful to introduce the magnetic Jeans length which allows to express the normalizations 
of the magnetic power spectra. The comoving magnetic Jeans length can be defined as \cite{bb0}
\begin{equation}
\lambda_{\mathrm{B\,J}} = c_{\mathrm{a}}^2 \sqrt{\frac{\pi}{G \, \overline{\rho}_{\mathrm{b}}}}, \qquad 
c_{\mathrm{a}}^2 = \frac{B_{\mathrm{L}}^2}{8 \pi \overline{\rho}_{\mathrm{b}}},
\label{MJ1}
\end{equation}
where $\overline{\rho}_{\mathrm{b}} = a^3 \rho_{\mathrm{b}}$ simply denote the comoving 
baryonic density. From Eq. (\ref{MJ1}) the explicit form of the square of the magnetic jeans length, 
becomes
\begin{equation}
\lambda_{\mathrm{B\,J}}^2 = \frac{8\pi^2}{3} \, \biggl(\frac{h_{0}}{H_{0}}\biggr)^2  \, 
\biggl(\frac{\omega_{\gamma\,0}}{\omega_{\mathrm{b}0}}\biggr) \overline{\Omega}_{\mathrm{B\,L}},
\label{MJ2}
\end{equation}
or, even more explicitly, 
\begin{equation}
\lambda_{\mathrm{B\,J}}  = 1.90\times 10^{-2} \, \biggl(\frac{\omega_{\mathrm{b}0}}{0.02258}\biggr)^{-1} \, 
\biggl(\frac{B_{\mathrm{L}}}{\mathrm{nG}}\biggr)\,\,\mathrm{Mpc}.
\label{MJ3}
\end{equation}
In terms of the magnetic Jeans wavenumber $k_{\mathrm{B\,J}} = 2\pi/\lambda_{\mathrm{B\,J}}$ 
the quantity $W_{\mathrm{B}}(k,n_{\mathrm{B}}, B_{\mathrm{L}})$ can be written as
\begin{equation}
W_{\mathrm{B}}(k,n_{\mathrm{B}}, B_{\mathrm{L}}) = \frac{\omega_{\mathrm{b}0}^2}{\omega_{\mathrm{M}0}}
\biggl(\frac{k}{k_{\mathrm{B\,J}}}\biggr)^2 \biggl(\frac{k}{k_{\mathrm{L}}}\biggr)^{n_{\mathrm{B}}-1} \, \biggl(\frac{k}{k_{\mathrm{p}}}\biggr)^{(1 - n_{\mathrm{s}})/2} \, \frac{{\mathcal L}(n_{\mathrm{B}}, k_{\mathrm{D}})}{\sqrt{{\mathcal A}_{{\mathcal R}}}\, \ln{(k/k_{\mathrm{eq}})}}.
\label{MJ4}
\end{equation}
The expression (\ref{MJ4}) generalizes to the $\Lambda$CDM situation the analog expressions arising 
in the absence of a dominant adiabatic mode seeding structure formation. Equation (\ref{MJ4}) 
shows that the effect of scale variation of the growth factor is partially dominated by scales 
close to the magnetic Jeans scale at least in the case $n_{\mathrm{s}} \simeq 1$ and $1< n_{\mathrm{B}} < 1.5$,
 which is what the plots of Figs. \ref{figure2} and \ref{figure3} show. 
\renewcommand{\theequation}{4.\arabic{equation}}
\setcounter{equation}{0}
\section{Numerical discussion} 
\label{sec4}
The results of section \ref{sec3} neglect the contribution of the radiation and of the relativistic corrections; furthermore 
they also assume that the evolution of the dark energy fluctuations does not appreciably contribute to the final shape of the growth factor.
In the absence of large-scale magnetic fields, the first two assumptions are rather reasonable for low redshifts and for modes which crossed inside 
the Hubble radius prior to equality, the third assumption is quantitatively correct only in $\Lambda$CDM context but 
not in the  $w$CDM case. If the plasma is magnetized the system changes qualitatively and quantitatively 
 both before and after photon decoupling. The numerical analysis of the present section is 
 intended to complement and corroborate the results of the previous sections.

The numerical integration of the system is carried on directly in terms of $y = \ln{\alpha}$ where,  we remind, 
$\alpha= a/a_{\mathrm{de}}$ and $\alpha_{\mathrm{de}}$ denotes the value of the scale factor 
when dark energy and matter give equal contribution to the total energy density. 
The variable $y$  is directly related to the redshift so that by plotting the growth factor in terms 
of $y$ also its redshift dependence can be easily sorted out since the following chain of equalities holds:
\begin{equation}
y = \ln{\alpha} = \ln{(z_{\mathrm{de}} +1)} - \ln{(z+1)}.
\label{Y1}
\end{equation}
The physical range of $y$ does not extend beyond $1$, indeed the dependence of $y_{\mathrm{max}}$ 
upon $w_{\mathrm{de}}$ is monotonic and ranges between $0.338$ (for $w_{\mathrm{de}} = -1$ and for the 
parameters of Eq. (\ref{FL11})) and $1.12$ in the case $w_{\mathrm{de}} = - 0.3$ (where, strictly speaking the 
background geometry does not accelerate). 
It is practical to introduce the rescaled wavenumber 
\begin{equation}
{\mathcal K}^2(y) = \frac{k^2}{{\mathcal H}^2} = \biggl(\frac{k}{H_{0}}\biggr)^2 \frac{(z_{\mathrm{eq}}+1)}{\Omega_{\mathrm{M}0} ( z_{\mathrm{de}}+1)} \frac{e^{(3 w_{\mathrm{de}} +1)y}}{[ (z_{\mathrm{eq}} +1) e^{3 w_{\mathrm{de}} y} + ( z_{\mathrm{de}} +1) e^{(3 w_{\mathrm{de}} -1)y} + z_{\mathrm{eq}} +1]},
\label{KK}
\end{equation}
so that, for instance,  Eq. (\ref{C2}) will be rewritten as 
\begin{equation}
\frac{d \delta_{\mathrm{c}}}{d y} = \frac{1}{2} \frac{d h}{d y} - {\mathcal K}(y) \overline{\theta}_{\mathrm{c}},\qquad 
\frac{d \overline{\theta}_{\mathrm{c}}}{d y} + \overline{\theta}_{\mathrm{c}} =0,
\label{DC}
\end{equation}
where, for a generic species $X$, $\overline{\theta}_{X} = \theta_{X}/k$. 
The evolution of $\xi(k,y)$ and $h(k,y)$ is determined, respectively, by the following pair of equations
\begin{eqnarray}
\frac{d^{2} \xi}{d y^2} + [ 2 + {\mathcal Z}(y)] \frac{d \xi}{dy}  &=& 2 \Omega_{\mathrm{R}} (R_{\nu}\sigma_{\nu} + R_{\gamma} \sigma_{\mathrm{B}})  - 
\frac{\Omega_{\mathrm{R}}}{2} ( \delta_{\mathrm{R}} + R_{\gamma} \Omega_{\mathrm{B}})  
\nonumber\\
&-& \frac{3}{2} \Omega_{\mathrm{de}} \biggl[ c_{\mathrm{de}}^2 \delta_{\mathrm{de}} + 
\frac{3}{{\mathcal K}(y)}(1 + w_{\mathrm{de}}) (c_{\mathrm{de}}^2 - w_{\mathrm{de}}) \overline{\theta}_{\mathrm{de}}\biggr],
\label{xi}\\
 \frac{d^{2} h}{d y^2} + [ 1 +{\mathcal Z}(y)]\frac{d h}{dy}  &=&
6 \, \Omega_{\mathrm{R}}[ \delta_{\mathrm{R}}  + R_{\gamma} \Omega_{\mathrm{B}}] 
+  3\, \Omega_{\mathrm{M}} \delta_{\mathrm{m}}
\nonumber\\
&+& 3 \Omega_{\mathrm{de}} \biggl[ (1+  3 c_{\mathrm{de}}^2)\delta_{\mathrm{de}} + 
\frac{9}{{\mathcal K}(y)} (1 + w_{\mathrm{de}})  (c_{\mathrm{de}}^2 - w_{\mathrm{de}}) 
\overline{\theta}_{\mathrm{de}}\biggr],
\label{h}
\end{eqnarray}
where ${\mathcal Z}(y)$ is expressible in terms of $y$ by recalling the definition of  Eq. (\ref{FL6})
and where the density contrasts in radiation and matter are defined as 
\begin{equation}
\delta_{\mathrm{R}}(k,y) = R_{\gamma} \delta_{\gamma}(k,y) + R_{\nu} \delta_{\nu}(k,y), \qquad 
\delta_{\mathrm{m}}(k,y) = \frac{\omega_{\mathrm{c}0}}{\omega_{\mathrm{M}0}} \delta_{\mathrm{c}}(k,y)
+ \frac{\omega_{\mathrm{b}0}}{\omega_{\mathrm{M}0}} \delta_{\mathrm{b}}(k,y).
\label{TOTDC}
\end{equation}
Equations (\ref{xi}) and (\ref{h}) are derived as linear combinations of Eqs. (\ref{ham1}), (\ref{ij1}) and (\ref{ij2}) written in the $y$ parametrization.  Moreover, in Eqs. (\ref{h}), (\ref{xi}) and (\ref{TOTDC}) 
$R_{\gamma} = 1 - R_{\nu}$ and 
\begin{equation}
R_{\nu} = \frac{\rho_{\nu}}{\rho_{\gamma} + \rho_{\nu}}= \frac{3 \times(7/8)\times (4/11)^{4/3}}{1 + 3 \times(7/8)\times (4/11)^{4/3}} =0.4052
\label{Rnu}
\end{equation}
where $3$ counts the massless neutrino families, $(7/8)$ stems from the Fermi-Dirac statistics and $(4/11)^{4/3}$ comes from the kinetic temperature of neutrinos. 
Within the same notations, the 
evolution for the neutrinos (see Eqs. (\ref{N2}) and (\ref{N3})) become:
\begin{eqnarray}
&& \frac{d \overline{\theta}_{\nu}}{d y} = \frac{{\mathcal K}(y)}{4} \delta_{\nu} - {\mathcal K}(y) \sigma_{\nu},\qquad 
\frac{d \delta_{\nu}}{d y} = \frac{2}{3} \frac{d h}{dy} - \frac{4}{3} {\mathcal K}(y) \overline{\theta}_{\nu}
\label{DN1}\\
&& \frac{ d \sigma_{\nu}}{d y} = \frac{4}{15} {\mathcal K}(y) \overline{\theta}_{\nu} - \frac{2}{15} \biggl[ \frac{d h}{dy} + 6 \frac{d \xi}{dy}\biggr].
\label{DN2}
\end{eqnarray}
Prior to recombination the tight-coupling limit has been enforced. To zeroth order in the tight coupling expansion 
the photon quadrupole as well as the polarization vanish and the
relevant evolution equations for the dipole and for the monopole will then become:
\begin{eqnarray}
&& \frac{d \overline{\theta}_{\gamma\mathrm{b}}}{d y} + \frac{R_{\mathrm{b}}}{R_{\mathrm{b}} +1 }
 \overline{\theta}_{\gamma\mathrm{b}} = \frac{{\mathcal K}(y) \delta_{\gamma}}{4 (R_{\mathrm{b}} + 1)} - 
 \frac{{\mathcal K}(y)}{ 4 [R_{\mathrm{b}}(y) + 1]} ( 4 \sigma_{\mathrm{B}} - \Omega_{\mathrm{B}})
 \label{TC1}\\
&& \frac{d \delta_{\mathrm{b}}}{d y} = \frac{1}{2} \frac{d h}{d y}  - {\mathcal K}(y)  \overline{\theta}_{\gamma \mathrm{b}},\qquad  \frac{d \delta_{\gamma}}{d y} = \frac{2}{3} \frac{ d h}{d y} - \frac{4}{3} {\mathcal K}(y) \overline{\theta}_{\gamma\mathrm{b}}.
\label{TC3}
\end{eqnarray}
After recombination the baryon and the photon velocities are different, i.e. 
\begin{equation}
(\overline{\theta}_{\mathrm{b}} - \overline{\theta}_{\gamma}) = \frac{ R_{\mathrm{b}}(y) \overline{{\mathcal K}}(y)}{[1 + R_{\mathrm{b}}(y)]{\mathcal K}(y)} \biggl\{ - \overline{\theta}_{\mathrm{b}}
+\frac{(\Omega_{\mathrm{B}} - 4 \sigma_{\mathrm{B}})}{4 R_{\mathrm{b}}(y)}   + 
{\mathcal K}(y) \biggl( c_{\mathrm{s\,b}}^2 \delta_{\mathrm{b}} - \frac{\delta_{\gamma}}{4} \biggr) + 
(\frac{d \overline{\theta}_{\gamma} }{d y} - \frac{d\overline{\theta}_{\mathrm{b}}}{d y})\biggr\},
\label{TC4}
\end{equation}
where the contribution of the quadrupole moment of the photon distribution has been neglected since it vanishes 
to zeroth-order in the tight coupling expansion; note that, as before, ${\mathcal K}(y)= k/{\mathcal H}$ while
$\overline{{\mathcal K}}(y) = k/\epsilon'$; $c_{\mathrm{s\, b}}^2$ denotes the sound speed in the baryon-photon 
fluid \cite{mg1,mg2}.  The temperature of the baryons has been taken to coincide with the temperature of the 
photons both before and after recombination.
After a transient regime the evolution equations for the photons and for the baryons will obey separate 
evolution equations. More specifically the appropriate collective variables will be the total 
density contrasts of matter and radiation as well as the radiation and matter velocities:
\begin{equation}
\overline{\theta}_{\mathrm{R}}(k,y) = R_{\gamma} \overline{\theta}_{\gamma}(k,y) + R_{\nu} \overline{\theta}_{\nu}(k,y), \qquad 
\overline{\theta}_{\mathrm{m}}(k,y) = \biggl(\frac{\omega_{\mathrm{c}0}}{\omega_{\mathrm{M}0}}\biggr) 
\overline{\theta}_{\mathrm{c}}(k,y)
+ \biggl(\frac{\omega_{\mathrm{b}0}}{\omega_{\mathrm{M}0}}\biggr) \overline{\theta}_{\mathrm{b}}(k,y).
\label{def}
\end{equation}
Finally the evolution equation for the dark energy (see Eqs. (\ref{DE2}) and (\ref{DE3})) can be written as
\begin{eqnarray}
&& \frac{ d \delta_{\mathrm{de}}}{d y} + 3 ( c_{\mathrm{de}}^2 - w _{\mathrm{de}}) \delta_{\mathrm{de}} + 
(w_{\mathrm{de}} + 1) {\mathcal K}(y) \overline{\theta}_{\mathrm{de}} 
\nonumber\\
&&+ 9 \frac{(1 + w_{\mathrm{de}})}{{\mathcal K}(y)}
 ( c_{\mathrm{de}}^2 - w _{\mathrm{de}}) \overline{\theta}_{\mathrm{de}} - \frac{ 1 + w_{\mathrm{de}}}{2} \frac{d h}{d y} =0
\label{TC5}\\
&& \frac{d \overline{\theta}_{\mathrm{de}}}{d y} + ( 1 - 3 c_{\mathrm{de}}^2) \overline{\theta}_{\mathrm{de}} + 
\frac{ c_{\mathrm{de}}^2 {\mathcal K}(y)}{ ( w_{\mathrm{de}} +1 )} \delta_{\mathrm{de}} =0.
\label{TC6}
\end{eqnarray}
In the case of adiabatic initial conditions the asymptotic solution of the previous set of equations can be written as 
\begin{eqnarray}
 \xi(k,y) &=& {\mathcal R}_{*}(k) + \biggl\{\frac{R_{\gamma} [ 4 \sigma_{\mathrm{B}}(k) - R_{\nu} \Omega_{\mathrm{B}}(k)]}{ 6 ( 4 R_{\nu} + 15)}  - \frac{4 R_{\nu} + 5}{12( 4 R_{\nu} + 15)} {\mathcal R}_{*}(k)  \biggr\} {\mathcal K}^2(y),
\label{AD1}\\
h(k,y) &=& \frac{{\mathcal R}_{*}(k)}{2} {\mathcal K}^2(y) - \frac{1}{36} \biggl\{- \frac{8 R_{\nu}^2 - 14 R_{\nu} - 75}{2(2 R_{\nu} + 25)(4 R_{\nu} + 15)} {\mathcal R}_{*}(k)
\nonumber\\
&+& \frac{R_{\gamma} ( 15 - 20 R_{\nu})}{10( 4 R_{\nu} + 15) ( 2 R_{\nu} + 25)} [R_{\nu}\Omega_{\mathrm{B}}(k) - 4 \sigma_{\mathrm{B}}(k)]\biggr\} {\mathcal K}^4(y),
\label{AD2}\\
\delta_{\gamma}(k,y) &=& - R_{\gamma} \Omega_{\mathrm{B}}(k) + \frac{2}{3} \biggl[ 
\frac{{\mathcal R}_{*}(k)}{2} + \sigma_{\mathrm{B}}(k) - \frac{R_{\nu}}{4} \Omega_{\mathrm{B}}(k)\biggr] {\mathcal K}^2(y),
\label{AD33}\\
\delta_{\nu}(k,y) &=& - R_{\gamma} \Omega_{\mathrm{B}}(k) + \frac{2}{3} \biggl[ \frac{{\mathcal R}_{*}(k)}{2}  - \frac{R_{\gamma}}{ R_{\nu}} \sigma_{\mathrm{B}}(k) - \frac{R_{\gamma}}{4} \Omega_{\mathrm{B}}(k)\biggr] {\mathcal K}^2(y),
\label{AD4}\\
\delta_{\mathrm{c}}(k,y) &=& - \frac{3}{4}R_{\gamma} \Omega_{\mathrm{B}}(k) + \frac{{\mathcal R}_{*}(k)}{4} {\mathcal K}^2(y),
\label{AD5}\\
\delta_{\mathrm{b}}(k,y) &=& - \frac{3}{4}R_{\gamma} \Omega_{\mathrm{B}}(k) + \frac{1}{2} \biggl[ \frac{{\mathcal R}_{*}(k)}{2}  + \sigma_{\mathrm{B}}(k)- \frac{R_{\nu}}{4} \Omega_{\mathrm{B}}(k) \biggr] {\mathcal K}^2(y),
\label{AD6}\\
\delta_{\mathrm{de}}(k,y) &=& - \frac{3}{4} ( 1 + w_{\mathrm{de}}) R_{\gamma} \Omega_{\mathrm{B}}(k) + \frac{(w_{\mathrm{de}}+1)}{4} {\mathcal R}_{*}(k) {\mathcal K}^2(y),
\label{AD6a}\\
\overline{\theta}_{\gamma\mathrm{b}}(k,y) &=& \biggl[ \frac{R_{\nu}}{4} \Omega_{\mathrm{B}}(k) - \sigma_{\mathrm{B}}\biggr] 
{\mathcal K}(y) -\frac{1}{36} \biggl[ -{\mathcal R}_{*}(k) + \frac{R_{\nu} \Omega_{\mathrm{B}}(k) - 4 \sigma_{\mathrm{B}}(k)}{2}\biggr] {\mathcal K}^3(y),
\label{AD7}\\
\overline{\theta}_{\nu}(k,y) &=& \biggl[ \frac{R_{\gamma}}{R_{\nu}} \sigma_{\mathrm{B}}(k) - \frac{R_{\gamma}}{4} \Omega_{\mathrm{B}}(k)\biggr] {\mathcal K}(y)
- \frac{1}{36}\biggl\{-\frac{( 4 R_{\nu} + 23)}{4 R_{\nu} + 15} {\mathcal R}_{*}(k) 
\nonumber\\
&+& \frac{R_{\gamma}( 4 R_{\nu} + 27)}{2 R_{\nu} ( 4 R_{\nu} + 15)}[ 4 \sigma_{\mathrm{B}}(k) - R_{\nu} \Omega_{\mathrm{B}}(k)]\biggr\} {\mathcal K}^3(y),
\label{AD8}\\
\overline{\theta}_{\mathrm{c}}(k,y) &=& 0,
\label{AD9}\\
\overline{\theta}_{\mathrm{de}}(k,y) &=& -  \frac{3 {\mathcal K}(y)c_{\mathrm{de}}^2}{4(2 - 3 c_{\mathrm{de}}^2)} R_{\gamma} \Omega_{\mathrm{B}}(k) + \frac{{\mathcal R}_{*}(k)}{4 ( 2 - 3 c_{\mathrm{de}}^2)} {\mathcal K}^3(y)
\label{AD10}\\
\sigma_{\nu}(k,y) &=& - \frac{R_{\gamma}}{R_{\nu}} \sigma_{\mathrm{B}}(k) + \biggl\{ -\frac{2 {\mathcal R}_{*}(k)}{3( 4 R_{\nu} + 15)} + \frac{R_{\gamma}[ 4 \sigma_{\mathrm{B}}(k) - R_{\nu} \Omega_{\mathrm{B}}(k)]}{ 2 R_{\nu}(4 R_{\nu} + 15)}\biggr\} {\mathcal K}^2(y).
\label{AD11}
\end{eqnarray}
\begin{figure}[!ht]
\centering
\includegraphics[height=6.5cm]{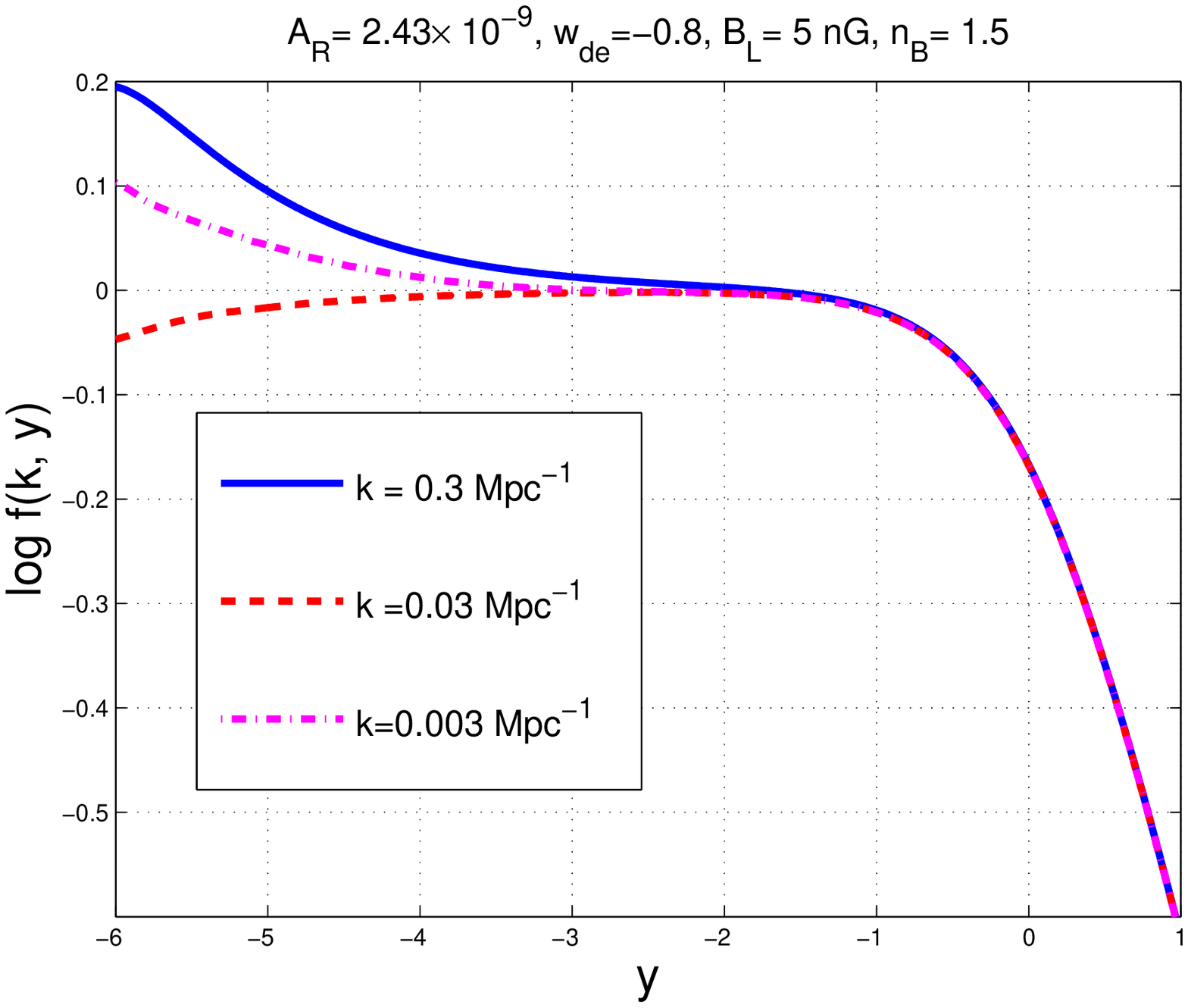}
\includegraphics[height=6.5cm]{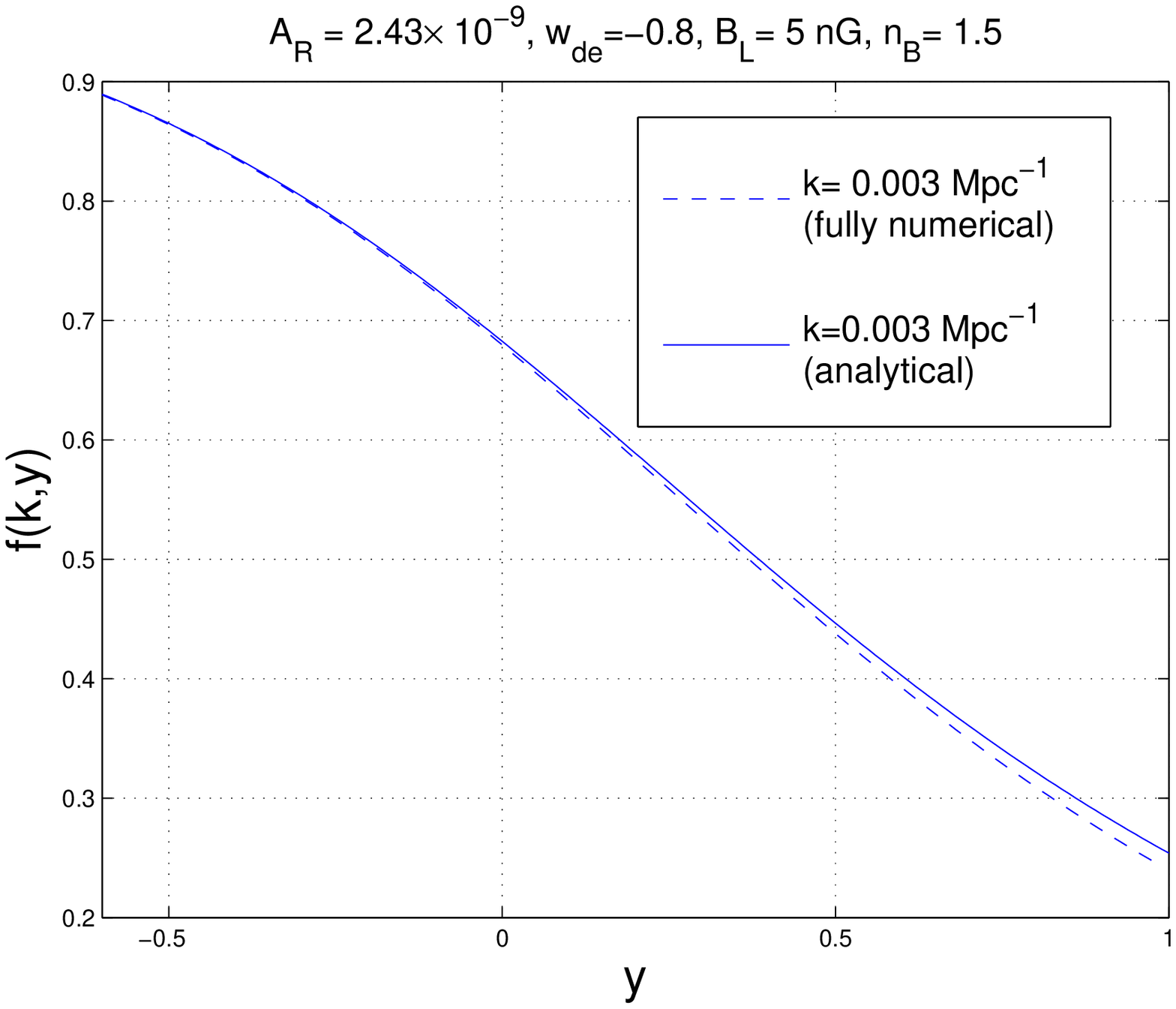}
\caption[a]{In the left plot the growth factor is illustrated for $w_{\mathrm{de}} =-0.8$
and in the case of three different wavenumbers, i.e. 
 $k= 0.3\, \mathrm{Mpc}^{-1}$ (full line), $k= 0.03\, \mathrm{Mpc}^{-1}$ (dashed line)
 and $k= 0.003\, \mathrm{Mpc}^{-1}$ (dot-dashed line). The right plot shows, for a particular 
 wavenumber, the result of the numerical integration of this section and the 
 analytical approximation based on Eq. (\ref{GR15}).}
\label{figure4}      
\end{figure}
As a cross-check of the accuracy of the integration the Hamiltonian and the momentum constraints 
must always be satisfied and their explicit expression becomes, respectively, 
\begin{eqnarray}
&& 2 {\mathcal K}(y) \xi - \frac{d h}{d y} - 3  [ \Omega_{\mathrm{R}} 
\delta_{\mathrm{R}} +  \Omega_{\mathrm{M}} \delta_{\mathrm{m}} + \Omega_{\mathrm{de}} 
\delta_{\mathrm{de}}]=0,
\label{hamex}\\
&& {\mathcal K}(y)\frac{d \xi}{d y} + 2 \Omega_{\mathrm{R}}(y) 
\overline{\theta}_{\mathrm{R}} + \frac{3}{2}( w_{\mathrm{de}} + 1) \Omega_{\mathrm{de}} 
\overline{\theta}_{\mathrm{de}} + \frac{3}{2}\Omega_{\mathrm{M}} \overline{\theta}_{\mathrm{m}}
=0.
\label{momex}
\end{eqnarray}
 The dimensionelss 
ratio defined in Eq. (\ref{KK}) dominates against all the terms containing explicitly 
the conductivity as implied by Eqs. (\ref{con1})--(\ref{con2}). 
To get an idea of the inaccuracies introduced by neglecting the diffusivity terms and the terms containing 
powers of the electric field a further dimensionless ratio can be defined and it is given 
 by ${\mathcal K}^2_{\sigma}(y) = k^2/({\mathcal H} \, \sigma)$. The ratio between ${\mathcal K}_{\sigma}(y)$ and ${\mathcal K}(y)$ 
gives 
\begin{equation}
\frac{{\mathcal K}_{\sigma}(y)}{{\mathcal K}(y)} < 4. 24 \times 10^{-13}  
\biggl(\frac{z_{\mathrm{drag}}}{1020.3}\biggr)^{-1/4} \biggl(\frac{h_{0}}{0.71}\biggr)^{1/2};
\label{con3}
\end{equation}
the inequality sign in Eq. (\ref{con3}) arises since we  assumed that in the dynamical range 
of the numerical integration $\Omega_{\mathrm{M}0} < \Omega_{\mathrm{M}}(y)$.  Thus 
the inaccuracies introduced by the resistive terms 
are of higher order in comparison with accuracy goal and with the working precision of the 
numerical calculation.
\begin{figure}[!ht]
\centering
\includegraphics[height=6.5cm]{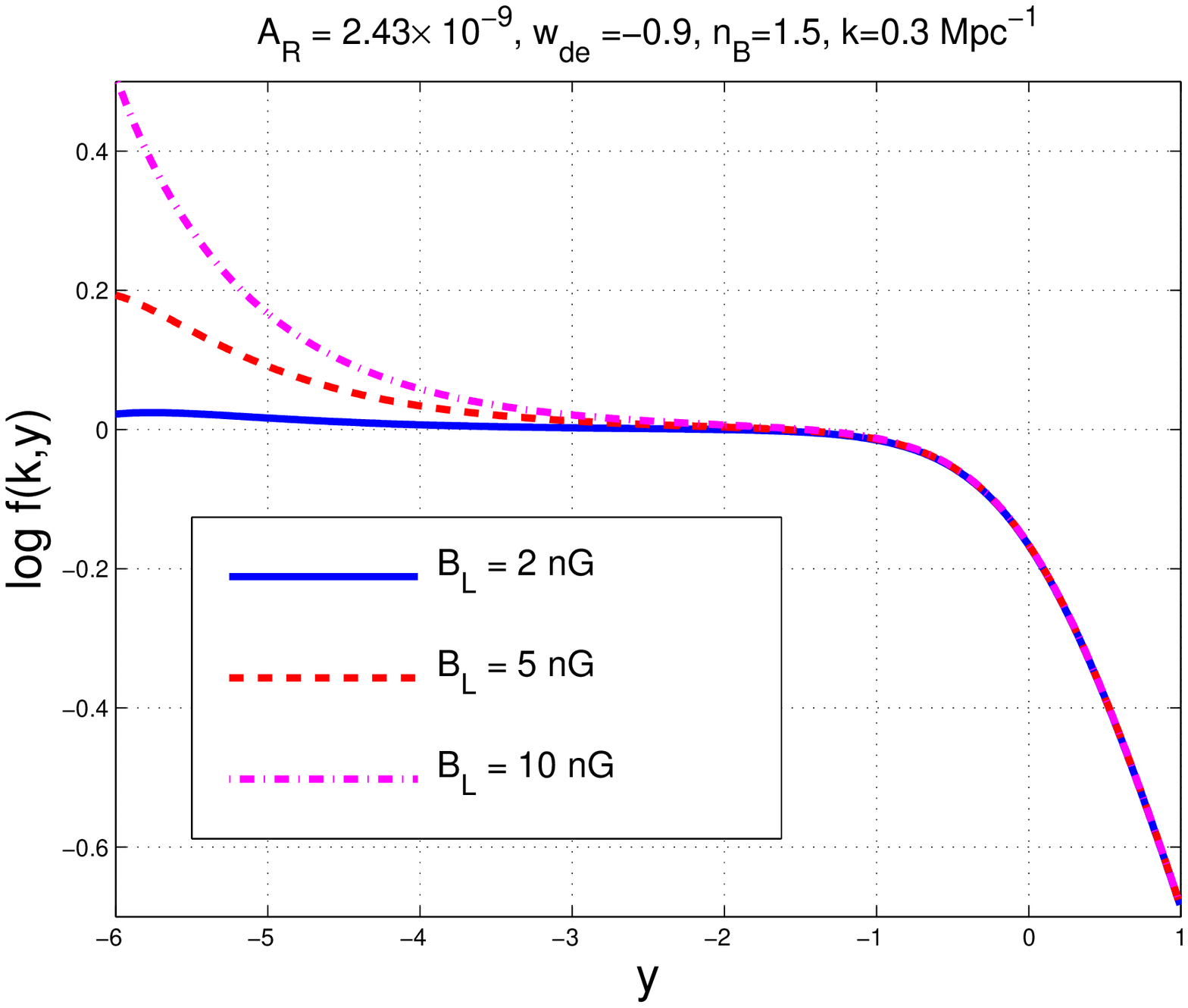}
\includegraphics[height=6.5cm]{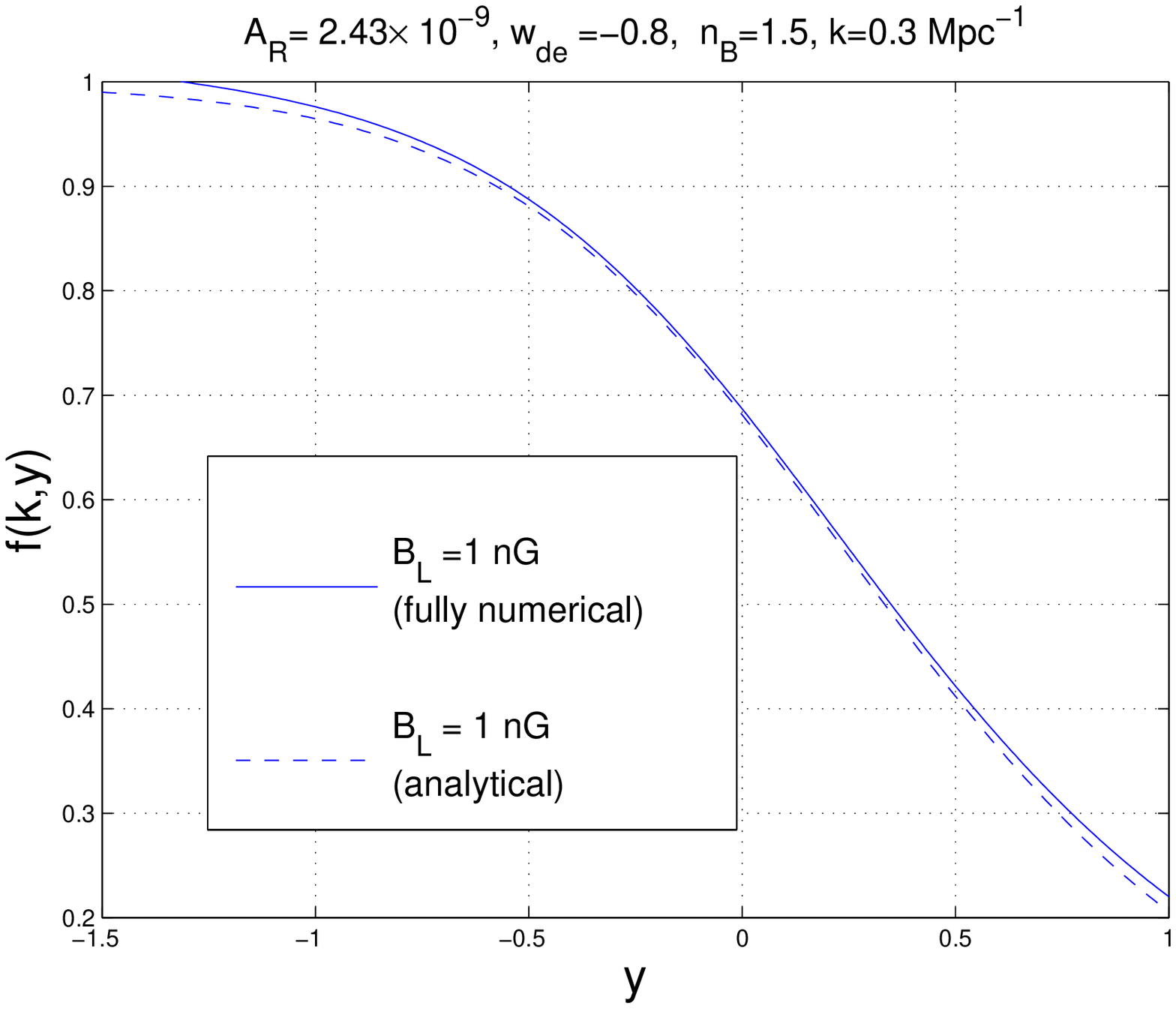}
\caption[a]{The growth factor is illustrated in the for different values of $B_{\mathrm{L}}$.}
\label{figure5}      
\end{figure}
The variation of the parameters affecting directly the shape of the magnetized growth factor will now be illustrated. 
The wavenumbers directly probed by the currently available large-scale 
structure data range from $k_{\mathrm{min}} = 0.01 \, h_{0}\, \mathrm{Mpc}^{-1}$ to
$k_{\mathrm{max}} = 0.25 \,h_{0}\, \mathrm{Mpc}^{-1}$ including also the typical range of scales at which the spectrum becomes nonlinear, i.e. $k_{\mathrm{nl}} \geq 0.09 \, h_{0} \, 
\mathrm{Mpc}^{-1}$. For illustration  three different values of $k$ are reported in the left plot of 
Fig. \ref{figure4}, i.e. $k= 0.3\, \mathrm{Mpc}^{-1}$ (full line), $k= 0.03\, \mathrm{Mpc}^{-1}$ (dashed line)
 and $k= 0.003\, \mathrm{Mpc}^{-1}$ (dot-dashed line).  Always in Fig. \ref{figure4} (plot at the right) the 
 numerical calculation described in the present section is compared with the analytical results of section \ref{sec3} for the same values of the parameters and by focusing the attention 
on the region of large $y$, i.e. $y \to 1$.  For the parameters chosen in Figs. \ref{figure4} and 
\ref{figure5}  the values of $y$ at decoupling and for different barotropic indices are given, respectively, by $y_{\mathrm{dec}} = - 6.569$ for 
$w= -0.8$, $y_{\mathrm{dec}} = - 6.616$  for $w=-0.9$ and $y_{\mathrm{dec}} = -6.653 $ for $w = -1$.
\begin{figure}[!ht]
\centering
\includegraphics[height=6.5cm]{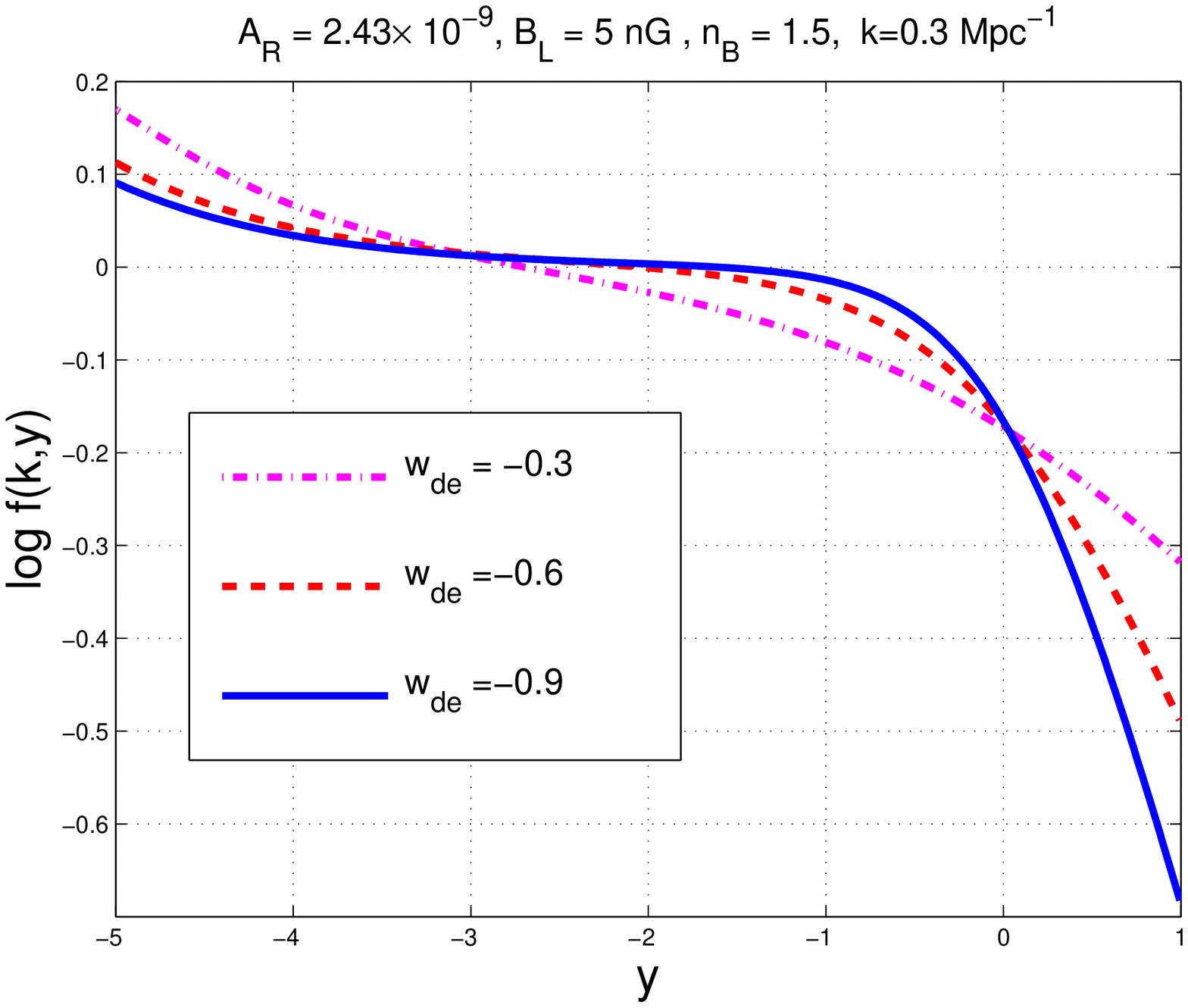}
\includegraphics[height=6.5cm]{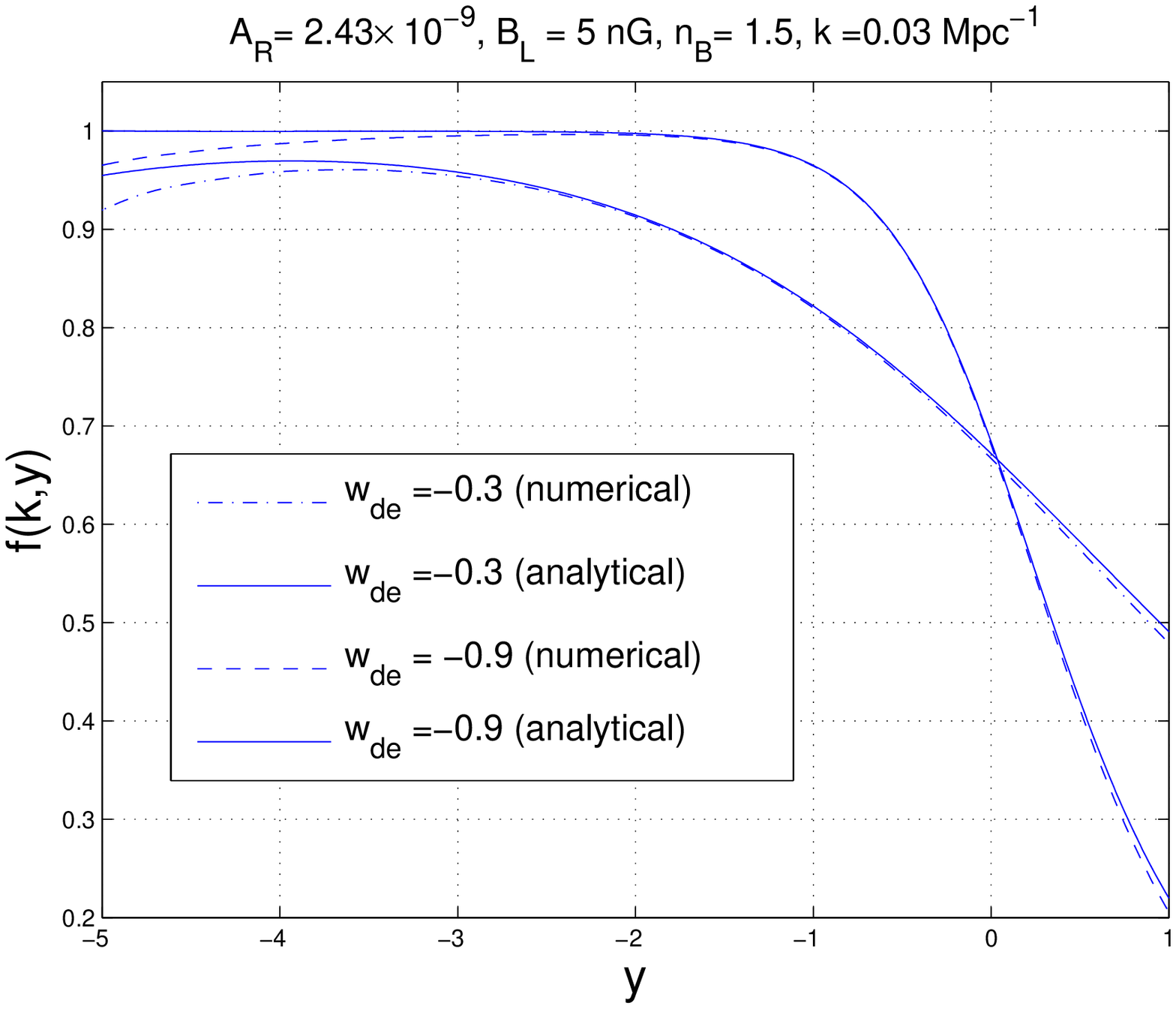}
\caption[a]{The growth rate is illustrated for different values of the barotropic index of the dark energy and compared with the analytical results of section \ref{sec3}.}
\label{figure6}      
\end{figure}
The parameters chosen for the numerical integrations of Figs. \ref{figure4} and \ref{figure5}
are the ones of Eq. (\ref{FL10}), where, however, $w_{\mathrm{de}}$ has been taken to be 
different from $-1$ to allow for dynamical dark energy fluctuations. The WMAP 7yr data 
alone are consistent with a rather broad interval of values of $w_{\mathrm{de}}$ ranging from 
$-0.63$ to about $-1.6$. We shall not contemplate the possibility of having $w_{\mathrm{de}} < -1$: in the 
latter case sudden singularities can arise in the future. In Figs. \ref{figure4}, \ref{figure5} and \ref{figure6} the speed of sound of dark energy is chosen to be $c_{\mathrm{de}}^2 =1$. The latter choice corresponds to 
identify the sound speed of dark energy with the speed of light as it happens, for instance, in quintessential 
models of dark energy. 
As already mentioned, the bounds on $c_{\mathrm{de}}^2$ are not tightly fixed by observation and the differences 
induced by different values of $c_{\mathrm{de}}^2$ will not be essential for the present discussion as long 
as $c_{\mathrm{de}}^2$ is positive semidefinite. A comparison between 
the case $c_{\mathrm{de}}^2 =1$ and $c_{\mathrm{de}}^2 = 0$ is drawn in Fig. \ref{figure6}. 
\begin{figure}[!ht]
\centering
\includegraphics[height=6.5cm]{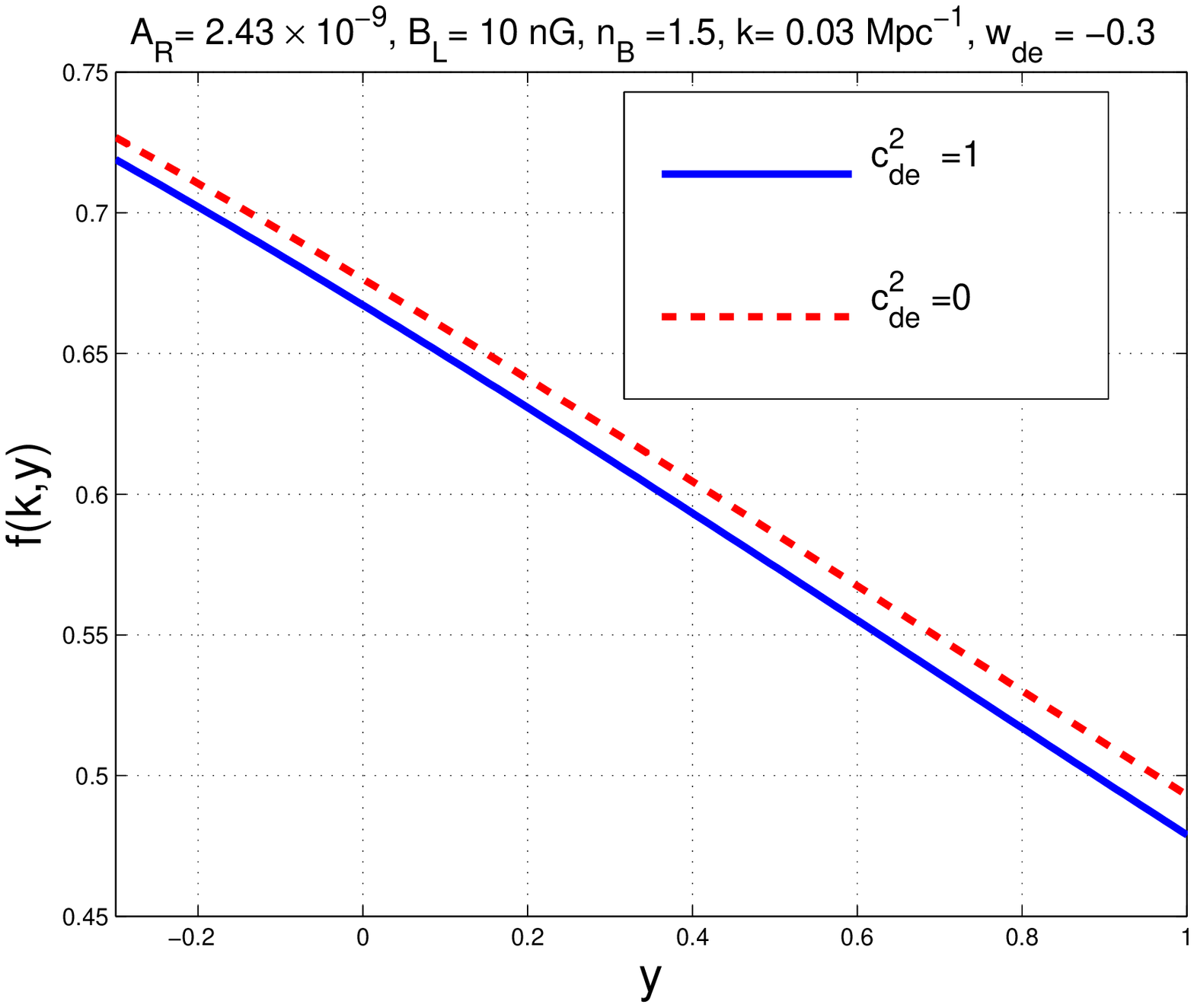}
\includegraphics[height=6.5cm]{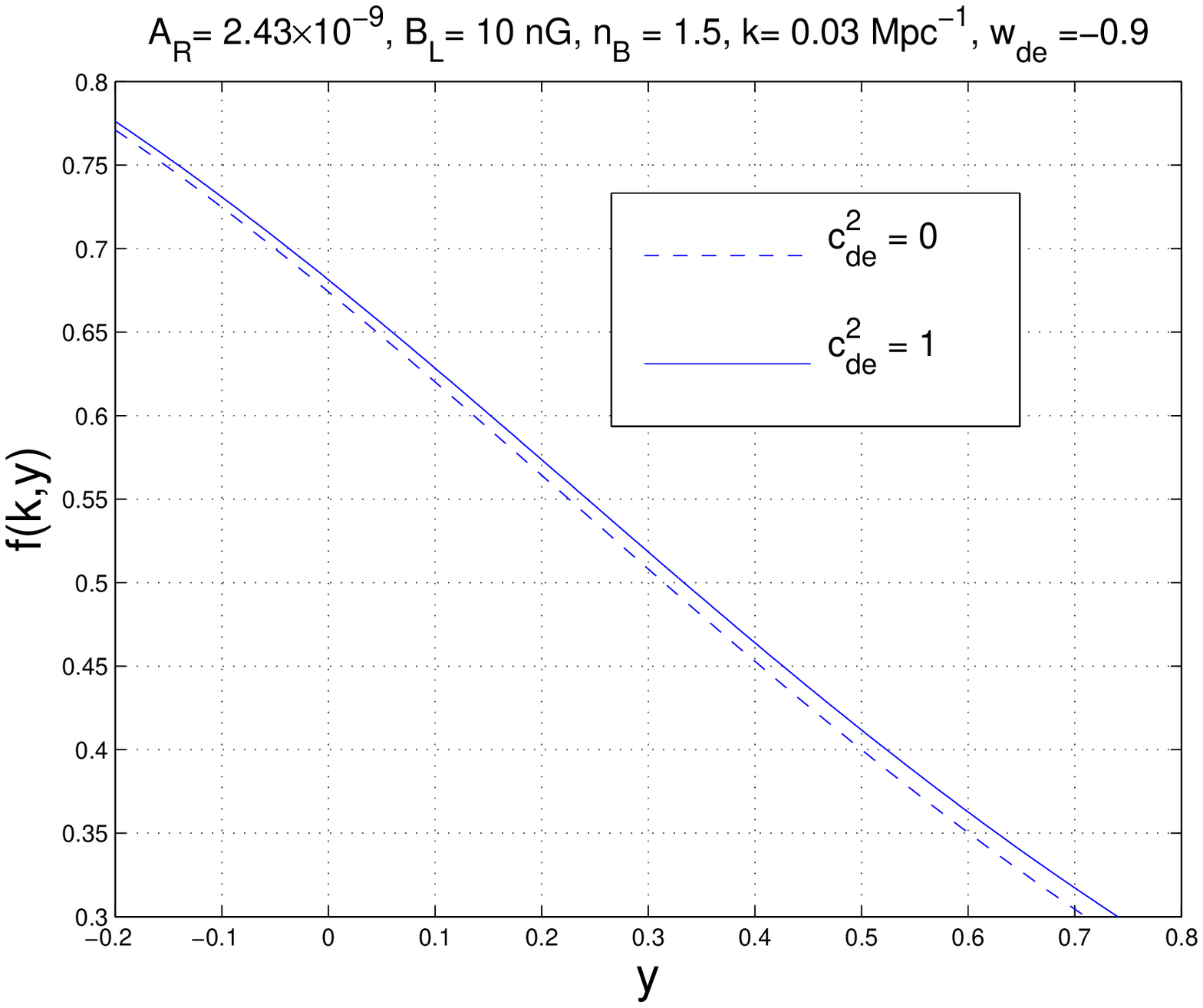}
\caption[a]{The variation of the sound speed of dark energy is illustrated for the two extreme values of the barotropic index adopted in the text. In the plot at the left $w_{\mathrm{de}} =-0.3$; in the plot at the right $w_{\mathrm{de}} =-0.9$.}
\label{figure7}      
\end{figure}
In Fig. \ref{figure5} the growth factor is illustrated for the variation of $B_{\mathrm{L}}$. The comparison 
of the numerical result with the analytical estimates is illustrated in the plot at the right. 
The differences between the numerical results and the analytical formula are determined, in this case, 
by the presence of dark energy fluctuations. Note that the range of $y$ has been narrowed 
to emphasize the differences. The same aspect is illustrated in Fig. \ref{figure6} 
where the growth factor is computed numerically in the case of different barotropic indices.
In both plots of Fig. \ref{figure6} the sound speed of the dark energy  fluctuations coincides with the speed of light 
(i.e. $c_{\mathrm{de}}^2 = 1$).  The results presented so far support the view that general 
relativistic corrections alter (from $5$ to $10$ \%) the growth rate from $z \simeq 100$ down to $z \simeq 0$ already for $k\simeq 0.003 \, \mathrm{Mpc}^{-1}$.  

In the left plot of Fig. \ref{figure6} the value of the comoving wavenumber is fixed to  $k =0.3\, \mathrm{Mpc}^{-1}$  
and the numerically computed gowth rates are illustrated for three different values of the barotropic indices of 
dark energy ranging between $-0.3$ up to $-0.9$. In the same figure (but in plot at the right) the numerical results 
are compared with the analytic approximations  based on Eqs. (\ref{Nw1}) and (\ref{GR15})--(\ref{GR15b}) for 
a comoving wavenumber  $k =0.03\, \mathrm{Mpc}^{-1}$. A general feature 
of the obtained numerical results, as exemplified by Fig. \ref{figure6} is a disagreement between 
the analytic and the numerical results for very large or very small redshifts. For small $y$ (i.e. large redshifts) 
the radiation must be properly treated (and it has been instead neglected in section \ref{sec3}).
Similarly, at low redshifts (i.e. large $y$) the fluctuations of the dark energy give a correction which has been 
neglected in the analytical discussion. In spite of the mentioned caveats, however, it seems 
that the analytical treatment correctly captures the region of intermediate redshifts and it 
is, overall, quite reasonable. 
While in the previous figures the sound speed of the dark energy has been taken to coincide 
with the speed of light, in Fig. \ref{figure7} the variation of the sound speed is illustrated.
In Fig. \ref{figure7} in the left and right plots the barotropic indices take the two extreme 
values $w_{\mathrm{de}} =-0.3$ and $w_{\mathrm{de}} =-0.9$. In each of the two plots 
two different values of the sound speed of the dark energy (i.e. $c_{\mathrm{de}}^2 =1$ 
and $c_{\mathrm{de}}^2 =0$) are illustrated. 
The range of redshifts has been narrowed to make the differences more visible. As expected 
the differences between the two cases are more evident in the region of  
low redshifts (i.e. large $y$), in agreement with the previous remarks. 
\renewcommand{\theequation}{5.\arabic{equation}}
\setcounter{equation}{0}
\section{Concluding remarks} 
\label{sec5}
Direct measurements of the growth rate of matter fluctuations can be used as a probe of large-scale magnetism. In the conventional $\Lambda$CDM paradigm (and in its extensions) the 
baryons fall into the dark matter potential wells so that their corresponding growth rate is determined primarily by the density contrast of cold dark matter. The latter statement must be partially revised in a magnetized environment where the resulting growth index is determined by the competition of the dark matter fluctuations and by the magnetic inhomogeneities entering the evolution equations of the baryons. The rich physical structure of the magnetized initial conditions 
of the Einstein-Boltzmann hierarchy has been explored in all phenomenologically 
interesting cases contemplating the standard adiabatic mode as well as 
the various isocurvature modes. The improved understanding of magnetized CMB anisotropies suggests a more thorough investigation of the effects of large-scale magnetization on the current paradigms of structure formation.  The parameters minimizing the distortions of the temperature and polarization angular power spectra  induce computable modifications on the shape of the growth rate. The magnetized growth rate has been scrutinized both for small and for large redshifts in comparison with the typical time-scale at which the baryons are freed from the drag of the photons. The results reported here pave the way for a more systematic scrutiny of the impact of large-scale magnetic fields on the current paradigm of structure formation based on  the the $\Lambda$CDM scenario and on its neighboring extensions.

\end{document}